%% file: main.tex
\documentclass[conference]{IEEEtran}
\IEEEoverridecommandlockouts
    
\usepackage{amsmath}
\usepackage{graphicx} 
\usepackage{epstopdf}
\usepackage{amssymb}
\usepackage{amsfonts}
\usepackage{amsthm}
\usepackage{cite}
\usepackage{bm,comment}
\usepackage{algorithm}
\usepackage{algpseudocode}

\usepackage{subeqnarray}
\usepackage{subfigure}
\usepackage{multicol}
\usepackage{multirow}
\usepackage{diagbox}
\usepackage{slashbox}
\usepackage{stfloats}
\usepackage{float}
\usepackage{color} 
\usepackage{cases}
\usepackage{enumitem}
\usepackage{xcolor}

\usepackage{pgfplots}
\usepackage{tikz}
\usetikzlibrary{calc}
\makeatletter
\newcommand{\gettikzxy}[3]{%
  \tikz@scan@one@point\pgfutil@firstofone#1\relax
  \edef#2{\the\pgf@x}%
  \edef#3{\the\pgf@y}%
}
\usetikzlibrary{spy,backgrounds}

\usepackage{lipsum}

\newtheorem{remark}{\bf{Remark}}

\usepackage{geometry}
\usepackage{acronym}
\input{Acronyms.tex}

\begin{document}
\bstctlcite{IEEEexample:BSTcontrol}
\setlength{\textfloatsep}{4pt}
\title{Privacy Preservation in Delay-Based Localization Systems: Artificial Noise or Artificial Multipath?}


\author{{Yuchen Zhang\IEEEauthorrefmark{1}, Hui Chen\IEEEauthorrefmark{2}, and Henk Wymeersch\IEEEauthorrefmark{2}}\\
\IEEEauthorrefmark{1}Electrical and Computer Engineering, King Abdullah University of Sciences and Technology, Saudi Arabia\\ 
\IEEEauthorrefmark{2}Department of Electrical Engineering, 
Chalmers University of Technology, Sweden 
\thanks{This work was supported, in part, by the Swedish Research Council (project 2023-03821) and Chalmers Area of Advance Transport. }
}

\maketitle

\begin{abstract}
Localization plays an increasingly pivotal role in 5G/6G systems, enabling various applications. This paper focuses on the privacy concerns associated with delay-based localization, where unauthorized base stations attempt to infer the location of the end user. We propose a method to disrupt localization at unauthorized nodes by injecting artificial components into the pilot signal, exploiting model mismatches inherent in these nodes. Specifically, we investigate the effectiveness of two techniques, namely artificial multipath (AM) and artificial noise (AN), in mitigating location leakage. By leveraging the misspecified Cramér-Rao bound framework, we evaluate the impact of these techniques on unauthorized localization performance. Our results demonstrate that pilot manipulation significantly degrades the accuracy of unauthorized localization while minimally affecting legitimate localization. Moreover, we find that the superiority of AM over AN varies depending on the specific scenario.






\end{abstract}
\begin{IEEEkeywords}
Secure localization, artificial path, artificial noise, misspecified Cr\'{a}mer-Rao bound.
\end{IEEEkeywords}

\IEEEpeerreviewmaketitle
\section{Introduction}

Localization is a fundamental component in 5G/6G systems, facilitating a variety of innovative applications such as collaborative robots and augmented reality~\cite{behravan2022positioning}. Compared with angle-based localization, delay-based methods have the advantage of cost-effectiveness as only a single antenna is required~\cite{dwivedi2021positioning}. Specifically, \ac{tdoa} can be used with either a downlink positioning reference signal or an uplink sounding reference signal, while multi-\ac{rtt} that utilize both reference signals can support \ac{toa}-based localization~\cite{tr38855}. A more recent technical report, TR 38.859, has studied \ac{tdoa} and \ac{rtt}-based positioning using sideline communications, substantially extending localization coverage~\cite{tr38859}. The adoption of large bandwidth signals enhances delay estimation resolution and the resolvability of multipath, making the system capable of dealing with localization tasks in \ac{nlos} scenarios~\cite{deng2020tdoa}.

While location-based services unlock significant new capabilities, they also introduce critical concerns regarding privacy issues, as information leakage to unauthorized entities can monitor private behavior without permission~\cite{shokri2012protecting}.
To address location leakage, various approaches have been introduced at the physical layer. In multi-antenna localization systems, techniques such as null-space beamforming and directional jamming have been proposed to degrade the performance of unauthorized localization by compromising the quality of the received signal\cite{stefano2020cl,stefano2022icc}. In model-free deep learning-based localization systems, adversarial machine learning has been employed to mitigate location leakage by introducing perturbations to the pilot signal\cite{studer2024twc}. However, these schemes typically rely on the availability of either the channel state information (CSI) or the utilized neural network at unauthorized nodes. 
Recently, a location privacy-preserving technique devoid of CSI was introduced in \cite{urbashi2023icassp}. This method primarily manipulates the pilot signal to generate \ac{am}, thereby disrupting \ac{tdoa} estimation and consequently impeding localization performance at unauthorized nodes, all without necessitating access to their CSI. Besides, the injection of \acp{am} has been demonstrated to be superior to emitting \ac{an}. 

\begin{figure}[t]
\centering
\includegraphics[width=0.7\linewidth]{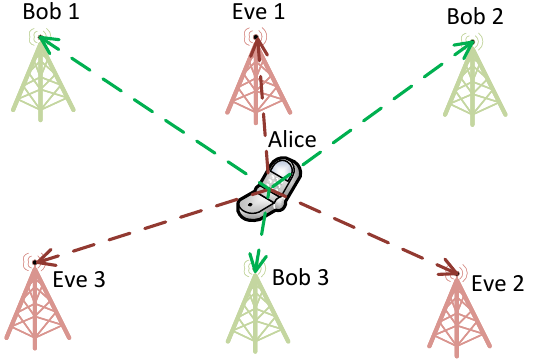}
\caption{Illustration of location leakage in a delay-based localization system (e.g., TDOA with 3 anchors, or TOA with 2 anchors up to an ambiguity). Alice modifies the uplink pilot to avoid being localized by the Eves.}\label{fig1}
\end{figure}

To quantify the performance of a localization system, the \ac{crb} is usually used. However, \ac{crb} fails to account for model mismatches caused by a modified pilot when the (unauthorized) \ac{bs} assumes a standard, pre-agreed signal. In such cases, instead of using \ac{crb} for privacy protection performance metrics~\cite{urbashi2023icassp}), the \ac{mcrb} is preferred~\cite{richmond2017spm}.
Previous studies using \ac{mcrb} have effectively analyzed various mismatch scenarios (e.g., using a far-field model in the near-field\cite{hui2022gc}, localization under hardware impairment~\cite{hui2023twc} and geometry error~\cite{pinjun2023icassp}, multipath scenarios~\cite{joseph2023taes}, and reconfigurable intelligent surface-aided systems~\cite{cuneyd2023twc,cuneyd2024twc}), demonstrating its utility in assessing the impact of mismatch factors.

In this work, we examine an uplink delay-based \ac{ofdm} localization system, reevaluating and comparing the effectiveness of \ac{an} and \ac{am} in protecting end user location information from unauthorized \acp{bs} by considering model mismatch (see Fig.~\ref{fig1}).
Our main contributions are summarized as follows: 
\begin{itemize}
\item We define a scenario in which an end device sends a pilot, altered by injecting artificial components, to induce erroneous position estimation at unauthorized \acp{bs}, exploiting the model mismatch present in unauthorized nodes;
\item Two strategies for mitigating location leakage are investigated (\ac{an} and \ac{am} injection), whose performance is systematically quantified through \ac{mcrb}-based analyses;
\item Numerical results show that pilot manipulation significantly degrades unauthorized localization performance, with minimal impact on legitimate localization. Additionally, \ac{am} does not consistently outperform \ac{an}, emphasizing the necessity of selecting location privacy-preserving techniques based on the specific situation.
\end{itemize}



\section{System Model}
As illustrated in Fig. \ref{fig1}, we consider an uplink \ac{tdoa}-based localization system in which several synchronized (legitimate) single-antenna \acp{bs} (Bobs) infer the location of single-antenna \ac{ue} (Alice) based on the delays estimated from the received uplink pilot. Due to its broadcasting nature, the pilot sent from Alice could also be eavesdropped by unauthorized single-antenna \acp{bs} (Eves), leading to potential threat of location leakage. 

\subsection{Signal Model}
Considering an \ac{ofdm} system with \ac{los} condition, Alice sends pilot $\mathbf{v}\in \mathbb{C}^{M \times 1}$ across $M$ subcarriers with a total bandwidth $W$. Here, $\|\mathbf{v}\|=\sqrt{P}$, where $P$ denotes the transmit power.
The signal received by a receiver (Bob or Eve) is given by
\begin{equation}\label{sig_mod}
\mathbf{y} = \alpha \mathbf{d}\left(\tau\right) \odot \mathbf{v} + \mathbf{n},
\end{equation}
where $\alpha$ is the complex channel gain, $\tau$ is the delay, $[\mathbf{d}\left(\tau\right)]_m=e^{-\jmath 2 \pi m \Delta f \tau}$ is the phase shifts across subcarriers, and $\mathbf{n} \sim \mathcal{CN}(\mathbf{0},N_0 \Delta f \mathbf{I}_M)$ is the additive Gaussian white noise (AWGN) with single-side \ac{psd} $N_0$. Essentially, if the pilot $\mathbf{v}$ is publicly known, Alice's location can be eavesdropped upon in a two-stage process utilizing the delay estimations from various Eves. Note that \ac{nlos} paths are not considered in the received signal, which is left for future work. Nevertheless, the proposed analysis can also be applied to cases with multipath, provided the \ac{los} is resolvable.

\subsection{Location Leakage Mitigation Strategies}

We describe two methods that Alice and Bob can employ to mitigate location leakage and thus preserve Alice’s privacy: \ac{an} and \ac{am} injection. In the next section, we will then discuss localization performance at Eve under mismatch caused by the \ac{an} and \ac{am}, to evaluate the performance of the privacy preservation schemes. For simplicity and without loss of generality, we consider $\mathbf{v} = \sqrt{P/M} \mathbf{1}_M$ and introduce $P_M=P/M$.

\subsubsection{Artificial Noise}
Under the \ac{an} strategy, the pilot is manipulated by integrating an AWGN-like perturbation as\cite{vknlau2017tvt} 
\begin{equation}\label{an}
    \mathbf{s}=\tilde{\gamma} \sqrt{P_M} \mathbf{1}_M+  \tilde{\gamma}\sqrt{\tilde{\beta}} \mathbf{z},
\end{equation}
where $\mathbf{z}\in \mathbb{C}^M$ is drawn from the standard complex normal distribution and then normalized to ensure $\left\| \mathbf{z} \right\| = \sqrt{P_M}$, $\tilde{\beta}$ characterizes the relative strength of the \ac{an} component, and $\tilde{\gamma}$ is a normalization factor to maintain $\left\|\mathbf{s}\right\| = \sqrt{P}$. Then, the signal received at a receiver is given by
\begin{equation}\label{rx_sig_an}
    \mathbf{y}= \alpha \tilde{\gamma} \sqrt{P_M} \mathbf{d}\left(\tau\right)+ \alpha \tilde{\gamma} \sqrt{\tilde{\beta}}  \mathbf{d}\left(\tau\right)\odot \mathbf{z}+ \mathbf{n}.
\end{equation}

\subsubsection{Artificial Multipath}
The concept of \ac{am} was introduced in\cite{urbashi2023icassp}, whose key idea is to manipulate the pilot by integrating perturbations, thereby creating \acp{am}. This compromises the delay estimations, and hence localization performance of unauthorized nodes who are unaware of the manipulation.
Specifically, the pilot is constructed as 
\begin{equation}\label{api}
    \mathbf{s}=\gamma \mathbf{v}+ \gamma\mathbf{v}\odot \sum_{l=1}^L \sqrt{\beta_l} \mathbf{d}\left(\delta_l\right),
\end{equation}
where $L$ is the number of artificial paths, $\beta_l$ characterizes the relative strength of the $l$-th component in the pilot, $\delta_l$ is the $l$-th differential delay, and $\gamma$ is a normalization factor to keep $\|\mathbf{s}\|=\sqrt{P}$. Similar to \cite{studer2024twc}, for the paths to be physically realizable with a time-domain filter, we consider that $\min_l \delta_l\ge0$ and $\max_l \delta_l < T_\text{CP}$, where $T_\text{CP}$ is the \ac{ofdm} \ac{cp} duration. 
The signal received at a receiver is expressed by 
\begin{equation}\label{rx_sig_simp}
    \mathbf{y}= \alpha \gamma \sum_{l=0}^L \sqrt{\beta_l} \sqrt{P_M} \mathbf{d}\left(\tau+\delta_l\right)+ \mathbf{n},
\end{equation}
where $\beta_0=1$ and $\delta_0=0$.

\section{Localization under Model Mismatch}
From the perspective of Eves, model mismatch occurs as the assumed pilot is $\mathbf{v}$ while the actual pilot is $\mathbf{s}$, resulting in a misspecified estimation problem, whose performance limit should be analyzed through the \ac{mcrb}\cite{richmond2017spm}. 

\subsection{MCRB Fundamentals}
Specifically, for a parameter $\boldsymbol{\theta}\in \mathbb{R}^{K\times 1}$, the \ac{lb} matrix for the mean squared error of a mismatched estimator is provided by\cite{richmond2017spm}
\begin{equation}\label{LBM}
\text{LB}\left(\boldsymbol{\theta},\boldsymbol{\theta}_0\right) =  \underbrace{\mathbf{A}^{-1}_{\boldsymbol{\theta}_0}\mathbf{B}_{\boldsymbol{\theta}_0}\mathbf{A}^{-1}_{\boldsymbol{\theta}_0}}_{\text{MCRB}\left(\boldsymbol{\theta}_0\right)}  + \underbrace{\left(\boldsymbol{\theta}-\boldsymbol{\theta}_0\right)\left(\boldsymbol{\theta}-\boldsymbol{\theta}_0\right)^\top}_{\text{Bias}\left(\boldsymbol{\theta}_0\right)}. 
\end{equation}
Here, $\boldsymbol{\theta}_0$ denotes the pseudo-true parameter obtained by\cite{richmond2017spm} 
\begin{equation}\label{pseud_para}
\boldsymbol{\theta}_0 = \arg \underset{\boldsymbol{\eta}}{\min\; }   \mathcal{D}\left(f_{\text{T}}\left(\mathbf{y}|\boldsymbol{\theta}\right)\left|\right|f_{\text{M}}\left(\mathbf{y}|\boldsymbol{\eta}\right)\right), 
\end{equation}
where $\boldsymbol{\eta}$ represents the parameter under mismatched model, and  $\mathcal{D}\left(f_{\text{T}}\left(\mathbf{y}|\boldsymbol{\theta}\right)\left|\right|f_{\text{M}}\left(\mathbf{y}|\boldsymbol{\eta}\right)\right) = \int_{f_T}{f_{\text{T}}\left(\mathbf{y}|\boldsymbol{\theta}\right)\ln \frac{f_{\text{T}}\left(\mathbf{y}|\boldsymbol{\theta}\right)}{f_{\text{M}}\left(\mathbf{y}|\boldsymbol{\eta}\right)}\mathrm{d}\mathbf{y}}$ denotes the Kullback-Leibler (KL) divergence between the true \ac{pdf} $f_{\text{T}}\left(\mathbf{y}|\boldsymbol{\theta}\right)$ and the mismatched \ac{pdf} $f_{\text{M}}\left(\mathbf{y}|\boldsymbol{\eta}\right)$.
Besides, $\mathbf{A}_{\boldsymbol{\theta}_0}$ and $\mathbf{B}_{\boldsymbol{\theta}_0}$ represent two generalizations of the Fisher information matrix (FIM), whose elements in the $i$-th row and the $j$-th column are determined by\cite{richmond2017spm}
\begin{eqnarray}\label{GFIM_A}
\begin{aligned}
\left[\mathbf{A}_{\boldsymbol{\theta}_0}\right]_{i,j}
=& 2 \text{Re}\left[\left(\frac{\partial^2 \boldsymbol{\mu}\left(\boldsymbol{\eta}\right)}{\partial\left[\boldsymbol{\eta}\right]_i \partial\left[\boldsymbol{\eta}\right]_j}\right)^{\mathsf{H}} \mathbf{C}_{\text{M}}^{-1}\boldsymbol{\epsilon}\left(\boldsymbol{\eta}\right)  \right.\\
-&\left.\left(\frac{\partial \boldsymbol{\mu}\left(\boldsymbol{\eta}\right)}{\partial\left[\boldsymbol{\eta}\right]_i }\right)^{\mathsf{H}}\mathbf{C}_{\text{M}}^{-1}\left(\frac{\partial \boldsymbol{\mu}\left(\boldsymbol{\eta}\right)}{ \partial\left[\boldsymbol{\eta}\right]_j}\right)\right]\bigg|_{\boldsymbol{\eta}=\boldsymbol{\theta}_0}
\end{aligned}
\end{eqnarray}
and 
\begin{align}\label{GFIM_B}
\left[\mathbf{B}_{\boldsymbol{\theta}_0}\right]_{i,j}
=& 4 \text{Re}\left[\boldsymbol{\epsilon}\left(\boldsymbol{\eta}\right)^{\mathsf{H}}\mathbf{C}_{\text{M}}^{-1}\frac{\partial \boldsymbol{\mu}\left(\boldsymbol{\eta}\right)}{\partial \left[\boldsymbol{\eta}\right]_i}\right]\text{Re}\left[\boldsymbol{\epsilon}\left(\boldsymbol{\eta}\right)^{\mathsf{H}} \mathbf{C}_{\text{M}}^{-1} \frac{\partial \boldsymbol{\mu}\left(\boldsymbol{\eta}\right)}{\partial \left[\boldsymbol{\eta}\right]_j}\right] \nonumber\\
+&2\text{Re}\left[\left(\frac{\partial \boldsymbol{\mu}\left(\boldsymbol{\eta}\right)}{\partial \left[\boldsymbol{\eta}\right]_i}\right)^{\mathsf{H}} \mathbf{C}_{\text{M}}^{-1} \left(\frac{\partial \boldsymbol{\mu}\left(\boldsymbol{\eta}\right)}{\partial \left[\boldsymbol{\eta}\right]_j}\right)\right]\bigg|_{\boldsymbol{\eta}=\boldsymbol{\theta}_0}
\end{align}
respectively. Here, $\mathbf{C}_{\text{M}}$ represents the covariance matrix of $f_{\text{M}}(\mathbf{y}|\boldsymbol{\eta})$, which is irrelevant to $\boldsymbol{\eta}$, and $\boldsymbol{\epsilon}(\boldsymbol{\eta}) = \boldsymbol{\kappa}(\boldsymbol{\theta}) - \boldsymbol{\mu}(\boldsymbol{\eta})$ with $\boldsymbol{\kappa}(\boldsymbol{\theta})$ and $\boldsymbol{\mu}(\boldsymbol{\eta})$ being the noise-free observations under the true and mismatched models, respectively\cite{pinjun2023icassp,hui2023twc}.

\newcounter{MYtempeqncn}
\begin{figure*}[!b]
\normalsize
\setcounter{MYtempeqncn}{\value{equation}}
\setcounter{equation}{10}
\hrulefill
\begin{equation}\label{mcrb_tau}
\text{MCRB}\left(\tau_0\right)=\frac{\left(6\left|\alpha\right| P \gamma \text{Im}\left[\xi\left(\tau_0\right)\right]  \right)^2 + 3M^2\left(M+1\right)\left(2M+1\right)N_0 \Delta f P}
{\left(12\pi \Delta f M^{-\frac{1}{2}}\left|\alpha\right| P^{\frac{3}{2}} \gamma  \text{Re}\left[\xi\left(\tau_0\right)\right] -6 \pi M^{\frac{1}{2}} \left(M+1\right) \Delta f \left|\alpha\right| P^{\frac{3}{2}} + 2\pi M\left(M+1\right)\left(2M+1\right) \Delta f \left|\alpha\right| P \right)^2}
\end{equation}
\setcounter{equation}{\value{MYtempeqncn}}
\end{figure*} 

\subsection{MCRB From Eve's Perspective}
Given that delays are independently estimated at various Eves before being combined to estimate Alice's location, determining the localization performance under model mismatch can proceed in two stages: 
\begin{enumerate}
    \item \emph{MCRB for Delay Estimation:}
The pseudo-true delays, along with their corresponding \ac{mcrb}s, are derived, forming another misspecified model regarding delay estimations. 
\item \emph{MCRB and LB for Location Estimation:} Leveraging the relationship between delays and location, the \ac{mcrb} and \ac{lb} of location estimation are determined. 
\end{enumerate}

We will make the assumption of \emph{powerful attackers}, where the only parameter each Eve  estimates is the delay $\tau$, while the complex channel gain $\alpha$ is known, and there is no clock offset between Alice and Eve. This assumption leads to a worst-case analysis from the perspective of legitimate nodes. The rationale behind this assumption is that if we can safeguard Alice's location from being leaked to Eves in the worst-case scenario, then more practical cases with weaker Eve  would not be worse. 

\subsubsection{\ac{mcrb} for Delay Estimation Under \ac{am}}

For the sake of notational convenience, the following derivation concerning delay estimation is performed at a specific Eve without specifying her index. Define noise-free received signals of the true model \eqref{rx_sig_simp} by $\boldsymbol{q}_{\text{T}}(\tau)=\alpha \gamma \sum_{l=0}^L \beta_l \sqrt{P_M} \mathbf{d}\left(\tau+\delta_l\right)$ and the mismatched model \eqref{sig_mod} by  $\boldsymbol{q}_{\text{M}}(\tau)=\alpha \sqrt{P_M} \mathbf{d}\left(\tau\right)$.
The true and mismatched \acp{pdf} of the received signal, conditioned on delay, are expressed as 
$f_{\text{T}}(\mathbf{y}|\tau)\propto \exp{(-{\left\|\mathbf{y}-\boldsymbol{q}_{\text{T}}(\tau)\right\|^2}/{(N_0 \Delta f)})}$ and $f_{\text{M}}(\mathbf{y}|\tau)\propto \exp{(-{\left\|\mathbf{y}-\boldsymbol{q}_{\text{M}}(\tau)\right\|^2}/{(N_0 \Delta f)})}$
%
By substituting these into \eqref{pseud_para}, the pseudo-true delay can be obtained by
\begin{align}
\tau_0 &= \arg \underset{\eta}{\min\; }   \mathcal{D}\left(f_{\text{T}}\left(\mathbf{y}|\tau\right)\left|\right|f_{\text{M}}\left(\mathbf{y}|\eta\right)\right) \notag
\\
&=\arg \underset{\eta}{\min\; }\left\|\boldsymbol{q}_{\text{T}}(\tau)-\boldsymbol{q}_{\text{M}}(\eta)\right\|^2
\notag
\\
&=\arg \underset{\eta}{\max\; }\sum_{m=1}^{M}\sum_{l=0}^{L}\sqrt{\beta_l}\cos \left(2\pi m \Delta f\left(\tau+\delta_l-\eta\right)\right).
\label{pseud_delay}
\end{align}
The above problem can be solved via line search. 
Then, through algebraic manipulation of \eqref{GFIM_A} and \eqref{GFIM_B}, the \ac{mcrb} regarding delay estimation is a scalar, as expressed in \eqref{mcrb_tau} at the bottom of this page, where 
\addtocounter{equation}{1}
\begin{align}
 \xi\left(\tau_0\right)=\sum_{m=1}^{M}\sum_{l=0}^{L} m \sqrt{\beta_l} \exp{\left(\jmath 2\pi m \Delta f\left(\tau+\delta_l-\tau_0 \right)\right)}.   
\end{align}
Note that the \ac{mcrb} in \eqref{mcrb_tau} degenerates to \ac{crb} (as will be introduced in \eqref{crb_mis_delay}) without model mismatch, i.e., when $\beta_l=0$ and $\delta_l=0\;(l=1,\ldots,L)$.

\subsubsection{\ac{mcrb} for Delay Estimation Under \ac{an}}
One can derive the \ac{mcrb} for delay estimation under \ac{an}, following a parallel process to that of its \ac{am} counterpart. For conciseness, we present the results directly. Specifically, the pseudo-true delay can be obtained by
\begin{eqnarray}\label{pseud_delay_AN}
\begin{aligned}
\tau_0 =\arg & \underset{\eta}{\max\; }\sum_{m=1}^{M}\left(\cos \left(2\pi m \Delta f\left(\tau-\eta\right)\right)\right.\\&
\left. +\sqrt{\tilde{\beta}}\left|z_m\right|\cos \left(2\pi m \Delta f\left(\tau-\eta\right) + \angle{z_m} \right) \right),
\end{aligned}
\end{eqnarray}
which can also be solved through line search. Moreover, the corresponding \ac{mcrb} is in the same form as \eqref{mcrb_tau}, albeit with 
\begin{align}
\xi(\tau_0) =\sum_{m=1}^{M} m (1+\sqrt{\tilde{\beta}}z_m)  \exp{\left(\jmath 2\pi m \Delta f(\tau-\tau_0 )\right)}. 
\end{align}

\begin{remark}
    For \ac{los} propagation, under both \ac{am} and \ac{an}, the pseudo-true delay from Eve $i$'s perspective can be expressed as $\tau_{0,i}=\tau_i + \Delta$, where $\Delta$ is independent of $i$. This is because the bias depends on the added perturbation in \eqref{an} and \eqref{api} is independent on Eve $i$.  For multipath scenario, however, $\Delta$ may no longer be identical for different Eves.
\end{remark}

\subsubsection{\ac{mcrb} and \ac{lb} for Location Estimation}
Based on the results pertaining to delay estimation under model mismatch, we can proceed to evaluate the \ac{mcrb} and \ac{lb} of location estimation, achieved through multiple cooperative Eves. Let $\boldsymbol{\tau}_{\text{E}}=[{\tau}_{\text{E},1},\ldots,{\tau}_{\text{E},K_{\text{E}}}]^\top$ and $\overline{\boldsymbol{\tau}}_{\text{E}}=[\overline{\tau}_{\text{E},1},\ldots,\overline{\tau}_{\text{E},K_{\text{E}}}]^\top$ represent the ground-truth delays and pseudo-true delays at the Eves, where $K_{\text{E}}$ denotes the number of Eves. Let $\mathbf{p}_{\text{A}}$ and $\mathbf{p}_{\text{E},i}$ denote the locations of Alice and the $i$-th Eve, respectively.\footnote{To maintain the generality, we do not explicitly specify the dimension of the location, as the derivations can be applied to both 2D and 3D cases.}  

Under the powerful attacker assumption, Eve is synchronized to Alice, so  $\tau_{\text{E},i} = \left\|\mathbf{p}_{\text{A}}-\mathbf{p}_{\text{E},i}\right\|/c$, where $c$ is the speed of light. The true and mismatched \acp{pdf} of the delay estimation $\hat{\boldsymbol{\tau}}_{\text{E}}$, conditioned on Alice's position, are expressed as
\begin{equation}\label{true_mod_pos}
f_{\text{T}}\left(\hat{\boldsymbol{\tau}}_{\text{E}}|\mathbf{p}_{\text{A}}\right)\propto  \exp{\left(-\frac{1}{2} \left(\hat{\boldsymbol{\tau}}_{\text{E}}-\overline{\boldsymbol{\tau}}_{\text{E}}\right)^\top \mathbf{\Xi}_{\text{T}}^{-1} \left(\hat{\boldsymbol{\tau}}_{\text{E}}-\overline{\boldsymbol{\tau}}_{\text{E}}\right)\right)}  
\end{equation}
and
\begin{equation}\label{mis_mod_pos}
f_{\text{M}}\left(\hat{\boldsymbol{\tau}}_{\text{E}}|\mathbf{p}_{\text{A}}\right)\propto  \exp{\left(-\frac{1}{2} \left(\hat{\boldsymbol{\tau}}_{\text{E}}-\boldsymbol{\tau}_{\text{E}}\right)^\top \mathbf{\Xi}_{\text{M}}^{-1}\left(\hat{\boldsymbol{\tau}}_{\text{E}}-\boldsymbol{\tau}_{\text{E}}\right)\right)} 
\end{equation}
respectively. Here, $\mathbf{\Xi}_{\text{T}} \in \mathbb{R}^{K_{\text{E}}\times K_{\text{E}}}$ and $\mathbf{\Xi}_{\text{M}} \in \mathbb{R}^{K_{\text{E}}\times K_{\text{E}}}$ are diagonal variance matrices with their $i$-th diagonal elements being $\text{MCRB}(\overline{\tau}_{\text{E},i})$ and $\text{CRB}({\tau}_{\text{E},i})$, respectively. In addition, $\text{CRB}({\tau}_{\text{E},i})$ denotes the \ac{crb} for delay estimation at the $i$-th Eve without model mismatch (i.e., in \eqref{sig_mod}), derived as 
\begin{equation}\label{crb_mis_delay}
\text{CRB}({\tau}_{\text{E},i})=\frac{3N_0}{4 \pi^2 \Delta f \left(M+1\right)\left(2M+1\right)\left|\alpha_{\text{E},i}\right|^2 P },    
\end{equation}
where $\alpha_{\text{E},i}$ is the complex channel gain at the $i$-th Eve.

We now proceed by deriving the pseudo-true location $\overline{\mathbf{p}}_{\text{A}}$, the MCRB, and the LB. 

\begin{itemize}[leftmargin=*]
    \item \emph{Pseudo-True Location:}
By substituting \eqref{true_mod_pos} and \eqref{mis_mod_pos} into \eqref{pseud_para}, the pseudo-true position of Alice can be obtained by
\begin{align}
\overline{\mathbf{p}}_{\text{A}} &= \arg \underset{\tilde{\mathbf{p}}_{\text{A}}}{\min\; }   \mathcal{D}\left(f_{\text{T}}\left(\hat{\boldsymbol{\tau}}_{\text{E}}|\mathbf{p}_{\text{A}}\right)\left|\right|f_{\text{M}}\left(\hat{\boldsymbol{\tau}}_{\text{E}}|\tilde{\mathbf{p}}_{\text{A}}\right)\right)\notag
\\
&=\arg \underset{\tilde{\mathbf{p}}_{\text{A}}}{\min\; } \left(\tilde{\boldsymbol{\tau}}_{\text{E}}(\tilde{\mathbf{p}}_{\text{A}})-\overline{\boldsymbol{\tau}}_{\text{E}}\right)^\top \mathbf{\Xi}_{\text{M}}^{-1} \left(\tilde{\boldsymbol{\tau}}_{\text{E}}(\tilde{\mathbf{p}}_{\text{A}})-\overline{\boldsymbol{\tau}}_{\text{E}}\right)   \notag
\\
&=\arg \underset{\tilde{\mathbf{p}}_{\text{A}}}{\min\; }\sum_{i=1}^{K_{\text{E}}}\frac{\left(\left\|\tilde{\mathbf{p}}_{\text{A}}-\mathbf{p}_{\text{E},i}  \right\|-c \overline{\tau}_{\text{E},i}  \right)^2}{\text{MCRB}\left(\overline{\tau}_{\text{E},i}\right)},
\label{pseud_pos}
\end{align}
where $\tilde{\boldsymbol{\tau}}_{\text{E}}=[\tilde{\tau}_{\text{E},1},\ldots,\tilde{\tau}_{\text{E},K_{\text{E}}}]^\top$ with $\tilde{\tau}_{\text{E},i} = \left\|\tilde{\mathbf{p}}_{\text{A}}-\mathbf{p}_{\text{E},i}\right\|/c$. The above problem can be solved using gradient descent with backtracking line search, wherein an initial point can be obtained via a coarse grid search\cite{cuneyd2023twc}.

\item \emph{MCRB:} To compute the \ac{mcrb} of location estimation, we need to determine the components therein. Specifically, for the first-order partial derivatives, we have
\begin{equation}\label{pos_1st_order_deriv}
\frac{\partial \boldsymbol{\tau}_{\text{E}} }{\partial \left[\mathbf{p}_{\text{A}}\right]_k } =\left[ \frac{\left[\mathbf{p}_{\text{A}}\right]_k-\left[\mathbf{p}_{\text{E},1}\right]_k}{c\left\| \mathbf{p}_{\text{A}}-\mathbf{p}_{\text{E},1}\right\|}, \ldots ,\frac{\left[\mathbf{p}_{\text{A}}\right]_k-\left[\mathbf{p}_{\text{E},K_{\text{E}}}\right]_k}{c\left\| \mathbf{p}_{\text{A}}-\mathbf{p}_{\text{E},K_{\text{E}}}\right\|} \right]^\top.
\end{equation}
For the second-order partial derivatives, we have
\begin{equation}\label{pos_2nd_order_deriv}
\frac{\partial^2 \boldsymbol{\tau}_{\text{E}} }{\partial \left[\mathbf{p}_{\text{A}}\right]_k \partial\left[\mathbf{p}_{\text{A}}\right]_n} =
\begin{cases}
 \boldsymbol{\psi}\left(\mathbf{p}_{\text{A}}\right), &    k=n,\\
 \mathbf{0}_{K_{\text{E}}}, & k \neq n,
\end{cases}
\end{equation}
where
\begin{equation}\label{pos_2nd_order_deriv_part2}
\boldsymbol{\psi}\left(\mathbf{p}_{\text{A}}\right)=\left[ \frac{1}{c\left\| \mathbf{p}_{\text{A}}-\mathbf{p}_{\text{E},1}\right\|}, \ldots ,\frac{1}{c\left\| \mathbf{p}_{\text{A}}-\mathbf{p}_{\text{E},K_{\text{E}}}\right\|} \right]^\top.
\end{equation}
Then, $\text{MCRB}\left(\overline{\mathbf{p}}_{\text{A}}\right)$ is obtained by substituting \eqref{pos_1st_order_deriv} and \eqref{pos_2nd_order_deriv}, along with $\mathbf{\Xi}_{\text{M}}$ and $\boldsymbol{\epsilon}(\tilde{\mathbf{p}}_{\text{A}}) =\tilde{\boldsymbol{\tau}}_{\text{E}}-\overline{\boldsymbol{\tau}}_{\text{E}}$, into \eqref{GFIM_A} and \eqref{GFIM_B}. 
\item \emph{LB:}
The \ac{lb} matrix is expressed as
\begin{equation}\label{LBM_pos}
\text{LB}\left(\mathbf{p}_{\text{A}},\overline{\mathbf{p}}_{\text{A}}\right) = \text{MCRB}\left(\overline{\mathbf{p}}_{\text{A}}\right)  + \left(\mathbf{p}_{\text{A}}-\overline{\mathbf{p}}_{\text{A}}\right)\left(\mathbf{p}_{\text{A}}-\overline{\mathbf{p}}_{\text{A}}\right)^\top. 
\end{equation}

Based on \eqref{LBM_pos}, the lower bound for the expected root mean squared error of position estimation in the presence of model mismatch is expressed as
\begin{equation}
\sqrt{\mathbb{E}\left[ \|\hat{\mathbf{p}}_{\text{A}}-\mathbf{p}_{\text{A}}\|^2  \right]} \ge \sqrt{\text{tr}\left(\text{LB}\left(\mathbf{p}_{\text{A}},\overline{\mathbf{p}}_{\text{A}}\right)\right)},   
\end{equation}
where $\hat{\mathbf{p}}_{\text{A}}$ denotes a misspecified-unbiased estimator, with its mean under the true model being $\overline{\mathbf{p}}_{\text{A}}$.
\end{itemize}

\subsection{Qualitative Analysis}

Based on the MCRB analysis, we  perform a qualitative performance prediction  in 2D  on the impact of both \ac{am} and \ac{an} as a function of the number of Eves. When there is only 1 Eve, \ac{am} and \ac{an} will constrain Eve's estimate of Alice on a circle around Eve, the radius of which depends on \ac{am} and \ac{an}. When there are 2 Eves, they will determine Alice on the intersection of two circles. This means that they will determine a location estimate (up to an ambiguity), with an error that depends on the LB. When there are 3 or more Eves, the localization problem becomes over-determined, which implies that methods such as TDOA can be applied. Given the observations in Remark 1, this implies that neither \ac{am} nor \ac{an} can protect Alice from the Eves determining her location.

{In the next section, we will use the derived bounds to quantitatively evaluate the impact of AM and AN on localization considering 2 Eves.}

\section{Numerical Results}
\subsection{Scenario}
Unless otherwise specified, the simulation parameters are presented as follows: A 2D localization scenario is considered where Alice is located at $[80\text{ m}, 80\text{ m}]^\top$, three Bobs (legitimate \acp{bs}) are located at $[0\text{ m}, 0\text{ m}]^\top$, $[90\text{ m}, 0\text{ m}]^\top$, and $[80\text{ m}, 160\text{ m}]^\top$, respectively, while two Eves (unauthorized \acp{bs}) are located at $[0\text{ m}, 0\text{ m}]^\top$ and $[80\text{ m}, 160\text{ m}]^\top$, respectively. 
The transmit power $P=10\text{ dBm}$, carrier frequency $f_c = 28 \text{ GHz}$, bandwidth $W = 100 \text{ MHz}$, number of subcarriers $M = 1024$, and noise \ac{psd} is $-173.855 \text{ dBm/Hz}$. In addition, the differential delays of \ac{am} follow $\delta_l = {l}/{(LW)}\; ( l=0,1,\ldots,L)$, with the maximum injected delay being the time resolution\footnote{If the maximum injected delay exceeds the time resolution, Eve can distinguish it as an additional path. Consequently, the mismatched model would not solely comprise a \ac{los} path, which is a case left for our future work.}, i.e., $1/W$.

\subsection{Results and Discussion}
We will first analyze the \ac{am} approach in detail, as it has more tunable parameters than the \ac{an} approach. Then, the impact of both \ac{an} and \ac{am} on localization will be evaluated under different scenarios.

\subsubsection{Impact of the Number of \acp{am}}
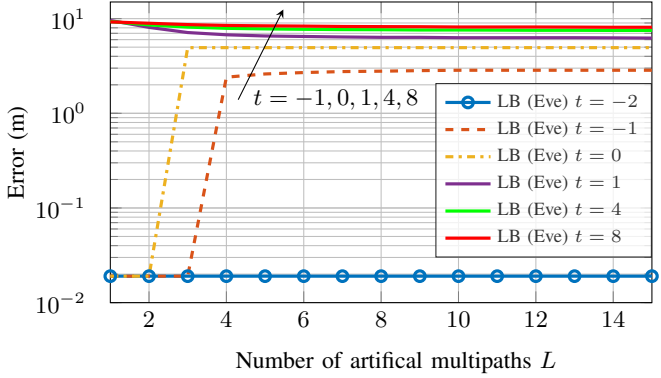
\begin{figure}[t]
\centering 
\centerline{\input{Figures/fig02}}
\caption{Location LB versus number of \acp{am} for various power allocation coefficients.}
\label{fig2}
\end{figure}

Fig. \ref{fig2} illustrates the effects of varying \ac{am} numbers $L$ on the \ac{lb} of unauthorized localization. To demonstrate the impact of the relative strength of the pilot's components, we consider $\beta_l = (l+1)^t\; ( l=0,1,\ldots,L)$ with $t$ as the decay factor. Specifically, for $t<0$, more power is allocated to the component with a smaller delay, for $t=0$, equal power is allocated to each component, and for $t>0$, less power is allocated to the component with a smaller delay. 
As can be concluded, for different $L$, allocating more power to the component with a larger delay, results in a larger \ac{lb} for unauthorized localization. Furthermore, keeping a minimal number of components intended for generating \acp{am} ($L=1$) is the most effective choice. The key insight here is that injecting more than one \ac{am} is not beneficial. Therefore, we set $L=1$ in the following simulations.

\subsubsection{Impact of \acp{am} Gain and Delay}
Fig. \ref{fig3} examines how different selections of $\beta_1$ and $\tilde{\delta}_1=\delta_1 W$ influence the \ac{lb} of unauthorized localization on a heat map. In this figure, when $\beta_1$ is sufficiently large, setting a larger $\tilde{\delta}_1$ (closer to 1) is more helpful in mitigating location leakage. This is because it increases the chance that Eve takes the generated \ac{am} with artificial delay as a LoS path, resulting in a larger bias in delay estimation and \ac{lb} of position estimation. 
An interesting phenomenon worth noting is that when $\beta_1$ is slightly less than $0 \text{ dB}$, i.e., the component in the pilot intended for generating the \ac{am} has almost the same power as the original component, the \ac{lb} of location estimation undergoes an up-then-down process as the injected delay becomes larger. 
This stems from the initial increase in injected delay, causing a larger bias in delay estimation and consequently larger \ac{lb} in position estimation due to insufficient separation between the two paths. As the injected delay gets close to the time resolution, the relatively stronger \ac{los} path is more distinguishable, resulting in reduced bias and \ac{lb} in position estimation. On the contrary, the \ac{nlos} dominates when $\beta_1 > 0 \text{ dB}$.

\begin{figure}[t]
\centering 
\centerline{\input{Figures/fig03}}
\caption{Heat map for localization LB versus $\beta_1$ and $\tilde{\delta}_1$.}
\label{fig3}
\end{figure}
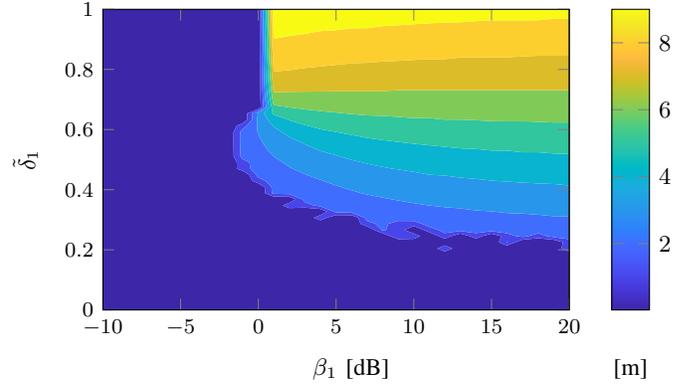


\begin{figure}[t]
\centering 
\centerline{\input{Figures/fig04}}
\caption{Localization error bounds (\ac{crb}, \ac{mcrb}, and \ac{lb}) and bias under \ac{am} and \ac{an}, versus $\beta_1$ or $\tilde{\beta}$, for $\delta_1=1/W$.}
\label{fig4}
\end{figure}
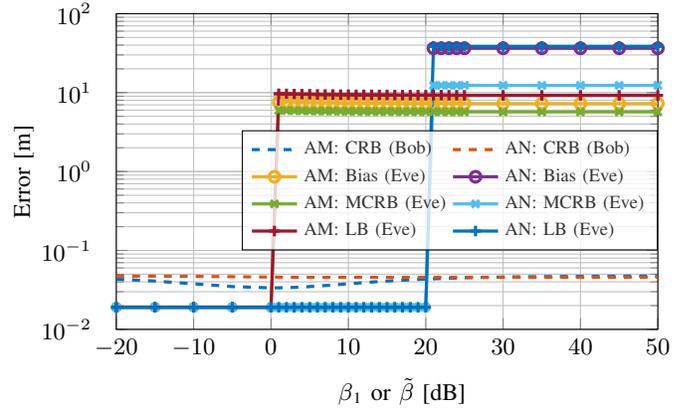



\begin{figure}[htb]
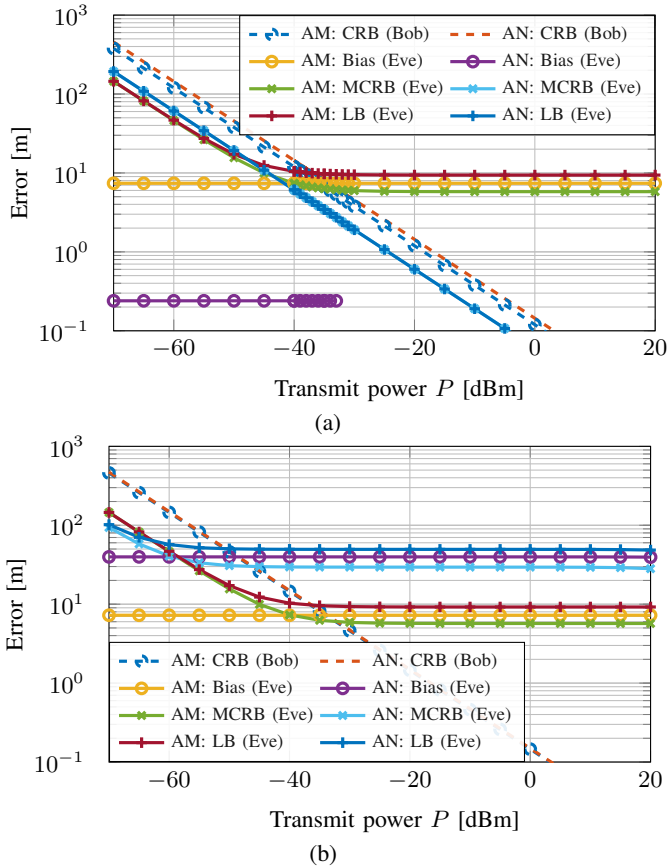

\centering
\begin{minipage}[b]{0.98\linewidth}
\vspace{-0.3cm}
  \centering
    \include{Figures/fig05a}
    \vspace{-1.cm}
  \centerline{\small{(a)}} \medskip
\end{minipage}
\hfill
\begin{minipage}[b]{0.98\linewidth}
\vspace{-0.3cm}
  \centering
    \include{Figures/fig05b}
    \vspace{-1.cm}
  \centerline{\small{(b)}} \medskip
\end{minipage}
\vspace{-0.2cm}
\caption{Localization error bounds (\ac{crb}, \ac{mcrb}, and \ac{lb}) and bias under \ac{am} and \ac{an}  for $\delta_1=1/W$, versus  
(a) Transmit power $P$ when $\beta_1=10\text{ dB}$.
(b) Transmit power $P$ when $\beta_1=30\text{ dB}$.}
\label{fig5}
\end{figure}

\subsubsection{Comparison of the Impact of the Injected Component}
In Fig. \ref{fig4}, we compare the \ac{lb} of both the legitimate localization at Bobs and the unauthorized localization at Eves under \ac{am} and \ac{an}. Specifically, we evaluate the impacts of the relative strength of injected components (indicated by $\beta_1$ under \ac{am} and $\tilde{\beta}$ under \ac{an}) on the localization performance. Note that the \ac{an} realization is fixed among different $\tilde{\beta}$ to remain consistence, and the position of sudden jump (i.e., $0 \text{ dB}$) may change with other realizations.
As seen, when the strength of the injected component is moderately larger than the original component, \ac{am} significantly outperforms \ac{an} in mitigating location leakage. However, when the injected component becomes more dominant, i.e., almost all power is allocated to it when formulating the pilot, \ac{an} exhibits superiority over \ac{am}. This results from the fact that the delay bias incurred by \ac{am} is limited by the maximum injected differential delay, which falls within the time resolution, while \ac{an} leads to a delay bias larger than this threshold. In the injected-component-dominant regime, the increased bias in delay estimation translates into a boost in \ac{lb} in position estimation. 

It is noteworthy that both \ac{am} and \ac{an} have an insignificant impact on legitimate localization in terms of \ac{crb}. This is because Bobs, as cooperative nodes, are aware of manipulation in the pilot without suffering from model mismatch. As long as the power of the pilot remains invariant, localization performance would not be severely impacted. However, it is observed that the \ac{crb} under \ac{am} is usually smaller than that under \ac{an}, especially when the injected and original components have relatively balanced power (e.g., $\beta_1 = 0$), demonstrating \ac{am}'s advantage in imposing less performance degradation towards legitimate localization.

\subsubsection{Comparison of the Impact of the Transmit Power}
Figs. \ref{fig5}(a) and (b) depict the localization error bounds and bias versus transmit power for $\beta_1\text{ or }\tilde{\beta}=10\text{ dB}$ and $\beta_1\text{ or }\tilde{\beta}=30\text{ dB}$, respectively. As observed, for unauthorized localization, the CRB decreases in the low-power regime and saturates as the power increases, demonstrating the mitigation of location leakage (i.e., the localization error is above a certain level). For legitimate localization without model mismatch, the CRB constantly decreases with increasing power. Note that we are analyzing the worst case where Even knows the clock offset and channel gain, and hence the LB is lower than CRB in the low-power regime where bias is not dominated. Additionally, when the transmit power is high (i.e., the noise power level is relatively low when $P>-30\text{ dBm}$), as shown in \ref{fig5}(a), AN will not introduce any bias term (only MCRB). Finally, as coincided with Fig. \ref{fig4}, \ac{am} is superior to \ac{an} with a moderately strong injected component (as shown in Fig. \ref{fig5}(a)), \ac{an} fails to introduce a large bias or enlarge variance) but less effective than \ac{an} when the injected component becomes more dominant (as shown in \ref{fig5}(b)).

\section{Conclusion}
This paper addressed the threat of location leakage in delay-based uplink localization systems, in which several unauthorized \acp{bs} could potentially infer the position of an end user. To protect the location privacy from being exposed to unauthorized \acp{bs}, we investigated two methods, namely \ac{am} and \ac{an}, whose key idea is manipulating the pilot by injecting an artificial component. This manipulation ensures that unauthorized \acp{bs}, without knowledge of the change in pilot, would undergo model mismatch and generate erroneous delay and location estimations. To analyze the performance of unauthorized localization, we resorted to the \ac{mcrb} analysis, tailored for evaluating estimation under model mismatch. Numerical results demonstrated that the manipulation in the pilot significantly undermined the performance of unauthorized localization while imposing marginal performance degradation to legitimate localization. Furthermore, the superiority of AM over AN varied depending on the specific scenario. Future work will extend the analytical framework to angle-based and scene-aware localization.

\appendices

\bibliographystyle{IEEEtran}
\bibliography{IEEEabrv,mybib}
\end{document}

%% file: Acronyms.tex
\acrodef{6g}[6G]{the sixth generation}
\acrodef{aoa}[AOA]{angle-of-arrival}
\acrodef{bs}[BS]{base station}
\acrodef{bse}[BSE]{beam squint effect}
\acrodef{cp}[CP]{cyclic prefix}
\acrodef{elaa}[ELAA]{extremely large antenna array}
\acrodef{ff}[FF]{far-field}
\acrodef{las}[L\&S]{localization and sensing}
\acrodef{los}[LOS]{line-of-sight}
\acrodef{nf}[NF]{near-field}
\acrodef{ris}[RIS]{reconfigurable intelligent surface}
\acrodef{rtt}[RTT]{round-trip-time}
\acrodef{sns}[SNS]{spatial non-stationarity}
\acrodef{swm}[SWM]{spherical wave model}

\acrodef{siso}[SISO]{single-input single-output}
\acrodef{ue}[UE]{user equipment}
\acrodef{dmimo}[D-MIMO]{distributed MIMO}

\acrodef{nlos}[NLOS]{non-line-of-sight}
\acrodef{ofdm}[OFDM]{orthogonal frequency division multiplexing}
\acrodef{tdoa}[TDOA]{time-difference-of-arrival}
\acrodef{toa}[TOA]{time-of-arrival}
\acrodef{am}[AM]{artificial multipath}
\acrodef{an}[AN]{artificial noise}
\acrodef{csi}[CSI]{channel state information}
\acrodef{mcrb}[MCRB]{misspecified Cramér-Rao bound}
\acrodef{crb}[CRB]{Cramér-Rao bound}
\acrodef{lb}[LB]{lower bound}
\acrodef{rmse}[RMSE]{root mean squared error}
\acrodef{psd}[PSD]{power spectral density}
\acrodef{pdf}[PDF]{probability distribution function}

%% file: Figures/fig02.tex
%
%
\definecolor{mycolor1}{rgb}{0.00000,0.44700,0.74100}%
\definecolor{mycolor2}{rgb}{0.85000,0.32500,0.09800}%
\definecolor{mycolor3}{rgb}{0.92900,0.69400,0.12500}%
\definecolor{mycolor4}{rgb}{0.49400,0.18400,0.55600}%
\begin{tikzpicture}

\begin{axis}[%
width=72mm,
height=40mm,
at={(0mm, 0mm)},
scale only axis,
xmin=1,
xmax=15,
xticklabel style = {font=\small,yshift=0ex},
xlabel style={font=\small, xshift=2mm},
xlabel={$\text{Number of artifical multipaths \it}L$},
ymode=log,
ymin=0.01,
ymax=15,
yminorticks=true,
yticklabel style = {font=\small,xshift=-2mm},
ylabel style={font=\small, yshift=0mm},
ylabel={Error (m)},
axis background/.style={fill=white},
xmajorgrids,
ymajorgrids,
yminorgrids,
legend style={font=\scriptsize, at={(1, 0.15)}, anchor=south east, legend cell align=left, align=left, draw=white!15!black, fill = white, fill opacity=0.8}
]
\addplot [color=mycolor1, line width=1.2pt, mark=o, mark options={solid, mycolor1}]
  table[row sep=crcr]{%
1	0.0190399733047829\\
2	0.0190399733047829\\
3	0.0190399733047829\\
4	0.0190399733047829\\
5	0.0190399733047829\\
6	0.0190399733047829\\
7	0.0190399733047829\\
8	0.0190399733047829\\
9	0.0190399733047829\\
10	0.0190399733047829\\
11	0.0190399733047829\\
12	0.0190399733047829\\
13	0.0190399733047829\\
14	0.0190399733047829\\
15	0.0190399733047829\\
};
\addlegendentry{LB (Eve) $t=-2$}

\addplot [color=mycolor2, line width=1.2pt,dashed]
  table[row sep=crcr]{%
1	0.0190399733047829\\
2	0.0190399733047829\\
3	0.0190399733047829\\
4	2.40965652543296\\
5	2.60442329682249\\
6	2.69552004865913\\
7	2.75987918721148\\
8	2.7863013104248\\
9	2.82408618504181\\
10	2.85041411567415\\
11	2.85041425253658\\
12	2.8504143122665\\
13	2.85041433183764\\
14	2.85041433447226\\
15	2.85041432693866\\
};
\addlegendentry{LB (Eve) $t=-1$}

\addplot [color=mycolor3, line width=1.2pt,dash dot]
  table[row sep=crcr]{%
1	0.0190399733047829\\
2	0.0190399733047829\\
3	4.93923942759183\\
4	4.93923942996241\\
5	4.93923943253931\\
6	4.93923943434166\\
7	4.93923943595093\\
8	4.93923943715\\
9	4.93923943813282\\
10	4.93923943897533\\
11	4.93923943965701\\
12	4.93923944027698\\
13	4.93923944071442\\
14	4.93923944126619\\
15	4.93923944152973\\
};
\addlegendentry{LB (Eve) $t=0$}

\addplot [color=mycolor4, line width=1.2pt]
  table[row sep=crcr]{%
1	9.57042860548872\\
2	8.08670487925348\\
3	7.13349430062006\\
4	6.75385409245844\\
5	6.56240103017814\\
6	6.48294600396732\\
7	6.40329912170454\\
8	6.34664461073535\\
9	6.34664455034973\\
10	6.2898953571105\\
11	6.28989535736592\\
12	6.28989535711998\\
13	6.28989535783377\\
14	6.26666270135367\\
15	6.23305193109709\\
};
\addlegendentry{LB (Eve) $t=1$}

\addplot [color=green, line width=1.2pt]
  table[row sep=crcr]{%
1	9.34131460826758\\
2	8.580553177548\\
3	8.14055732595584\\
4	7.9025387313467\\
5	7.79405031523755\\
6	7.71742010898489\\
7	7.66293373864522\\
8	7.64063071685551\\
9	7.58601990911969\\
10	7.58602352606057\\
11	7.56368055648124\\
12	7.53132768859918\\
13	7.53132258374244\\
14	7.5089448458983\\
15	7.50894732001429\\
};
\addlegendentry{LB (Eve) $t=4$}

\addplot [color=red, line width=1.2pt]
  table[row sep=crcr]{%
1	9.24646139930179\\
2	8.94130591059079\\
3	8.65538576046885\\
4	8.47406248367344\\
5	8.39886577965133\\
6	8.34539505604775\\
7	8.26994044395185\\
8	8.23823389763647\\
9	8.21627974744666\\
10	8.1625448145452\\
11	8.16254478245715\\
12	8.16254404108906\\
13	8.14055702318747\\
14	8.10873461200165\\
15	8.10873840679476\\
};
\addlegendentry{LB (Eve) $t=8$}

\end{axis}

\draw [-stealth] (1.7, 2.7) -- (2.3, 3.9);

\node[rotate=0] at (3.cm, 2.7cm) {\small{\shortstack[l]{$t = -1, 0, 1, 4, 8$}}};

\end{tikzpicture}%

%% file: Figures/fig03.tex
%
%
\begin{tikzpicture}

\begin{axis}[%
width=6.2cm,
height=4cm,
scale only axis,
scale only axis,
point meta min=0.0190399733047829,
point meta max=9,
axis on top,
xmin=-10,
xmax=20,
xticklabel style = {font=\footnotesize,yshift=0ex},
xlabel style={font=\small, xshift=2mm, yshift=0mm},
xlabel={$\beta_1$ [dB]},
ymin=0,
ymax=1,
yticklabel style = {font=\footnotesize,xshift=0 mm},
ylabel style={font=\small, yshift= -2mm},
ylabel={$\tilde{\delta}_1$},
axis background/.style={fill=white},
colormap={mymap}{[1pt] rgb(0pt)=(0.2422,0.1504,0.6603); rgb(1pt)=(0.2444,0.1534,0.6728); rgb(2pt)=(0.2464,0.1569,0.6847); rgb(3pt)=(0.2484,0.1607,0.6961); rgb(4pt)=(0.2503,0.1648,0.7071); rgb(5pt)=(0.2522,0.1689,0.7179); rgb(6pt)=(0.254,0.1732,0.7286); rgb(7pt)=(0.2558,0.1773,0.7393); rgb(8pt)=(0.2576,0.1814,0.7501); rgb(9pt)=(0.2594,0.1854,0.761); rgb(11pt)=(0.2628,0.1932,0.7828); rgb(12pt)=(0.2645,0.1972,0.7937); rgb(13pt)=(0.2661,0.2011,0.8043); rgb(14pt)=(0.2676,0.2052,0.8148); rgb(15pt)=(0.2691,0.2094,0.8249); rgb(16pt)=(0.2704,0.2138,0.8346); rgb(17pt)=(0.2717,0.2184,0.8439); rgb(18pt)=(0.2729,0.2231,0.8528); rgb(19pt)=(0.274,0.228,0.8612); rgb(20pt)=(0.2749,0.233,0.8692); rgb(21pt)=(0.2758,0.2382,0.8767); rgb(22pt)=(0.2766,0.2435,0.884); rgb(23pt)=(0.2774,0.2489,0.8908); rgb(24pt)=(0.2781,0.2543,0.8973); rgb(25pt)=(0.2788,0.2598,0.9035); rgb(26pt)=(0.2794,0.2653,0.9094); rgb(27pt)=(0.2798,0.2708,0.915); rgb(28pt)=(0.2802,0.2764,0.9204); rgb(29pt)=(0.2806,0.2819,0.9255); rgb(30pt)=(0.2809,0.2875,0.9305); rgb(31pt)=(0.2811,0.293,0.9352); rgb(32pt)=(0.2813,0.2985,0.9397); rgb(33pt)=(0.2814,0.304,0.9441); rgb(34pt)=(0.2814,0.3095,0.9483); rgb(35pt)=(0.2813,0.315,0.9524); rgb(36pt)=(0.2811,0.3204,0.9563); rgb(37pt)=(0.2809,0.3259,0.96); rgb(38pt)=(0.2807,0.3313,0.9636); rgb(39pt)=(0.2803,0.3367,0.967); rgb(40pt)=(0.2798,0.3421,0.9702); rgb(41pt)=(0.2791,0.3475,0.9733); rgb(42pt)=(0.2784,0.3529,0.9763); rgb(43pt)=(0.2776,0.3583,0.9791); rgb(44pt)=(0.2766,0.3638,0.9817); rgb(45pt)=(0.2754,0.3693,0.984); rgb(46pt)=(0.2741,0.3748,0.9862); rgb(47pt)=(0.2726,0.3804,0.9881); rgb(48pt)=(0.271,0.386,0.9898); rgb(49pt)=(0.2691,0.3916,0.9912); rgb(50pt)=(0.267,0.3973,0.9924); rgb(51pt)=(0.2647,0.403,0.9935); rgb(52pt)=(0.2621,0.4088,0.9946); rgb(53pt)=(0.2591,0.4145,0.9955); rgb(54pt)=(0.2556,0.4203,0.9965); rgb(55pt)=(0.2517,0.4261,0.9974); rgb(56pt)=(0.2473,0.4319,0.9983); rgb(57pt)=(0.2424,0.4378,0.9991); rgb(58pt)=(0.2369,0.4437,0.9996); rgb(59pt)=(0.2311,0.4497,0.9995); rgb(60pt)=(0.225,0.4559,0.9985); rgb(61pt)=(0.2189,0.462,0.9968); rgb(62pt)=(0.2128,0.4682,0.9948); rgb(63pt)=(0.2066,0.4743,0.9926); rgb(64pt)=(0.2006,0.4803,0.9906); rgb(65pt)=(0.195,0.4861,0.9887); rgb(66pt)=(0.1903,0.4919,0.9867); rgb(67pt)=(0.1869,0.4975,0.9844); rgb(68pt)=(0.1847,0.503,0.9819); rgb(69pt)=(0.1831,0.5084,0.9793); rgb(70pt)=(0.1818,0.5138,0.9766); rgb(71pt)=(0.1806,0.5191,0.9738); rgb(72pt)=(0.1795,0.5244,0.9709); rgb(73pt)=(0.1785,0.5296,0.9677); rgb(74pt)=(0.1778,0.5349,0.9641); rgb(75pt)=(0.1773,0.5401,0.9602); rgb(76pt)=(0.1768,0.5452,0.956); rgb(77pt)=(0.1764,0.5504,0.9516); rgb(78pt)=(0.1755,0.5554,0.9473); rgb(79pt)=(0.174,0.5605,0.9432); rgb(80pt)=(0.1716,0.5655,0.9393); rgb(81pt)=(0.1686,0.5705,0.9357); rgb(82pt)=(0.1649,0.5755,0.9323); rgb(83pt)=(0.161,0.5805,0.9289); rgb(84pt)=(0.1573,0.5854,0.9254); rgb(85pt)=(0.154,0.5902,0.9218); rgb(86pt)=(0.1513,0.595,0.9182); rgb(87pt)=(0.1492,0.5997,0.9147); rgb(88pt)=(0.1475,0.6043,0.9113); rgb(89pt)=(0.1461,0.6089,0.908); rgb(90pt)=(0.1446,0.6135,0.905); rgb(91pt)=(0.1429,0.618,0.9022); rgb(92pt)=(0.1408,0.6226,0.8998); rgb(93pt)=(0.1383,0.6272,0.8975); rgb(94pt)=(0.1354,0.6317,0.8953); rgb(95pt)=(0.1321,0.6363,0.8932); rgb(96pt)=(0.1288,0.6408,0.891); rgb(97pt)=(0.1253,0.6453,0.8887); rgb(98pt)=(0.1219,0.6497,0.8862); rgb(99pt)=(0.1185,0.6541,0.8834); rgb(100pt)=(0.1152,0.6584,0.8804); rgb(101pt)=(0.1119,0.6627,0.877); rgb(102pt)=(0.1085,0.6669,0.8734); rgb(103pt)=(0.1048,0.671,0.8695); rgb(104pt)=(0.1009,0.675,0.8653); rgb(105pt)=(0.0964,0.6789,0.8609); rgb(106pt)=(0.0914,0.6828,0.8562); rgb(107pt)=(0.0855,0.6865,0.8513); rgb(108pt)=(0.0789,0.6902,0.8462); rgb(109pt)=(0.0713,0.6938,0.8409); rgb(110pt)=(0.0628,0.6972,0.8355); rgb(111pt)=(0.0535,0.7006,0.8299); rgb(112pt)=(0.0433,0.7039,0.8242); rgb(113pt)=(0.0328,0.7071,0.8183); rgb(114pt)=(0.0234,0.7103,0.8124); rgb(115pt)=(0.0155,0.7133,0.8064); rgb(116pt)=(0.0091,0.7163,0.8003); rgb(117pt)=(0.0046,0.7192,0.7941); rgb(118pt)=(0.0019,0.722,0.7878); rgb(119pt)=(0.0009,0.7248,0.7815); rgb(120pt)=(0.0018,0.7275,0.7752); rgb(121pt)=(0.0046,0.7301,0.7688); rgb(122pt)=(0.0094,0.7327,0.7623); rgb(123pt)=(0.0162,0.7352,0.7558); rgb(124pt)=(0.0253,0.7376,0.7492); rgb(125pt)=(0.0369,0.74,0.7426); rgb(126pt)=(0.0504,0.7423,0.7359); rgb(127pt)=(0.0638,0.7446,0.7292); rgb(128pt)=(0.077,0.7468,0.7224); rgb(129pt)=(0.0899,0.7489,0.7156); rgb(130pt)=(0.1023,0.751,0.7088); rgb(131pt)=(0.1141,0.7531,0.7019); rgb(132pt)=(0.1252,0.7552,0.695); rgb(133pt)=(0.1354,0.7572,0.6881); rgb(134pt)=(0.1448,0.7593,0.6812); rgb(135pt)=(0.1532,0.7614,0.6741); rgb(136pt)=(0.1609,0.7635,0.6671); rgb(137pt)=(0.1678,0.7656,0.6599); rgb(138pt)=(0.1741,0.7678,0.6527); rgb(139pt)=(0.1799,0.7699,0.6454); rgb(140pt)=(0.1853,0.7721,0.6379); rgb(141pt)=(0.1905,0.7743,0.6303); rgb(142pt)=(0.1954,0.7765,0.6225); rgb(143pt)=(0.2003,0.7787,0.6146); rgb(144pt)=(0.2061,0.7808,0.6065); rgb(145pt)=(0.2118,0.7828,0.5983); rgb(146pt)=(0.2178,0.7849,0.5899); rgb(147pt)=(0.2244,0.7869,0.5813); rgb(148pt)=(0.2318,0.7887,0.5725); rgb(149pt)=(0.2401,0.7905,0.5636); rgb(150pt)=(0.2491,0.7922,0.5546); rgb(151pt)=(0.2589,0.7937,0.5454); rgb(152pt)=(0.2695,0.7951,0.536); rgb(153pt)=(0.2809,0.7964,0.5266); rgb(154pt)=(0.2929,0.7975,0.517); rgb(155pt)=(0.3052,0.7985,0.5074); rgb(156pt)=(0.3176,0.7994,0.4975); rgb(157pt)=(0.3301,0.8002,0.4876); rgb(158pt)=(0.3424,0.8009,0.4774); rgb(159pt)=(0.3548,0.8016,0.4669); rgb(160pt)=(0.3671,0.8021,0.4563); rgb(161pt)=(0.3795,0.8026,0.4454); rgb(162pt)=(0.3921,0.8029,0.4344); rgb(163pt)=(0.405,0.8031,0.4233); rgb(164pt)=(0.4184,0.803,0.4122); rgb(165pt)=(0.4322,0.8028,0.4013); rgb(166pt)=(0.4463,0.8024,0.3904); rgb(167pt)=(0.4608,0.8018,0.3797); rgb(168pt)=(0.4753,0.8011,0.3691); rgb(169pt)=(0.4899,0.8002,0.3586); rgb(170pt)=(0.5044,0.7993,0.348); rgb(171pt)=(0.5187,0.7982,0.3374); rgb(172pt)=(0.5329,0.797,0.3267); rgb(173pt)=(0.547,0.7957,0.3159); rgb(175pt)=(0.5748,0.7929,0.2941); rgb(176pt)=(0.5886,0.7913,0.2833); rgb(177pt)=(0.6024,0.7896,0.2726); rgb(178pt)=(0.6161,0.7878,0.2622); rgb(179pt)=(0.6297,0.7859,0.2521); rgb(180pt)=(0.6433,0.7839,0.2423); rgb(181pt)=(0.6567,0.7818,0.2329); rgb(182pt)=(0.6701,0.7796,0.2239); rgb(183pt)=(0.6833,0.7773,0.2155); rgb(184pt)=(0.6963,0.775,0.2075); rgb(185pt)=(0.7091,0.7727,0.1998); rgb(186pt)=(0.7218,0.7703,0.1924); rgb(187pt)=(0.7344,0.7679,0.1852); rgb(188pt)=(0.7468,0.7654,0.1782); rgb(189pt)=(0.759,0.7629,0.1717); rgb(190pt)=(0.771,0.7604,0.1658); rgb(191pt)=(0.7829,0.7579,0.1608); rgb(192pt)=(0.7945,0.7554,0.157); rgb(193pt)=(0.806,0.7529,0.1546); rgb(194pt)=(0.8172,0.7505,0.1535); rgb(195pt)=(0.8281,0.7481,0.1536); rgb(196pt)=(0.8389,0.7457,0.1546); rgb(197pt)=(0.8495,0.7435,0.1564); rgb(198pt)=(0.86,0.7413,0.1587); rgb(199pt)=(0.8703,0.7392,0.1615); rgb(200pt)=(0.8804,0.7372,0.165); rgb(201pt)=(0.8903,0.7353,0.1695); rgb(202pt)=(0.9,0.7336,0.1749); rgb(203pt)=(0.9093,0.7321,0.1815); rgb(204pt)=(0.9184,0.7308,0.189); rgb(205pt)=(0.9272,0.7298,0.1973); rgb(206pt)=(0.9357,0.729,0.2061); rgb(207pt)=(0.944,0.7285,0.2151); rgb(208pt)=(0.9523,0.7284,0.2237); rgb(209pt)=(0.9606,0.7285,0.2312); rgb(210pt)=(0.9689,0.7292,0.2373); rgb(211pt)=(0.977,0.7304,0.2418); rgb(212pt)=(0.9842,0.733,0.2446); rgb(213pt)=(0.99,0.7365,0.2429); rgb(214pt)=(0.9946,0.7407,0.2394); rgb(215pt)=(0.9966,0.7458,0.2351); rgb(216pt)=(0.9971,0.7513,0.2309); rgb(217pt)=(0.9972,0.7569,0.2267); rgb(218pt)=(0.9971,0.7626,0.2224); rgb(219pt)=(0.9969,0.7683,0.2181); rgb(220pt)=(0.9966,0.774,0.2138); rgb(221pt)=(0.9962,0.7798,0.2095); rgb(222pt)=(0.9957,0.7856,0.2053); rgb(223pt)=(0.9949,0.7915,0.2012); rgb(224pt)=(0.9938,0.7974,0.1974); rgb(225pt)=(0.9923,0.8034,0.1939); rgb(226pt)=(0.9906,0.8095,0.1906); rgb(227pt)=(0.9885,0.8156,0.1875); rgb(228pt)=(0.9861,0.8218,0.1846); rgb(229pt)=(0.9835,0.828,0.1817); rgb(230pt)=(0.9807,0.8342,0.1787); rgb(231pt)=(0.9778,0.8404,0.1757); rgb(232pt)=(0.9748,0.8467,0.1726); rgb(233pt)=(0.972,0.8529,0.1695); rgb(234pt)=(0.9694,0.8591,0.1665); rgb(235pt)=(0.9671,0.8654,0.1636); rgb(236pt)=(0.9651,0.8716,0.1608); rgb(237pt)=(0.9634,0.8778,0.1582); rgb(238pt)=(0.9619,0.884,0.1557); rgb(239pt)=(0.9608,0.8902,0.1532); rgb(240pt)=(0.9601,0.8963,0.1507); rgb(241pt)=(0.9596,0.9023,0.148); rgb(242pt)=(0.9595,0.9084,0.145); rgb(243pt)=(0.9597,0.9143,0.1418); rgb(244pt)=(0.9601,0.9203,0.1382); rgb(245pt)=(0.9608,0.9262,0.1344); rgb(246pt)=(0.9618,0.932,0.1304); rgb(247pt)=(0.9629,0.9379,0.1261); rgb(248pt)=(0.9642,0.9437,0.1216); rgb(249pt)=(0.9657,0.9494,0.1168); rgb(250pt)=(0.9674,0.9552,0.1116); rgb(251pt)=(0.9692,0.9609,0.1061); rgb(252pt)=(0.9711,0.9667,0.1001); rgb(253pt)=(0.973,0.9724,0.0938); rgb(254pt)=(0.9749,0.9782,0.0872); rgb(255pt)=(0.9769,0.9839,0.0805)},
colorbar,
colorbar style={font=\small, xshift=0mm, yshift=0mm, 
xlabel=[m]}
]
\addplot [forget plot] graphics [xmin=-10, xmax=20, ymin=0, ymax=1] {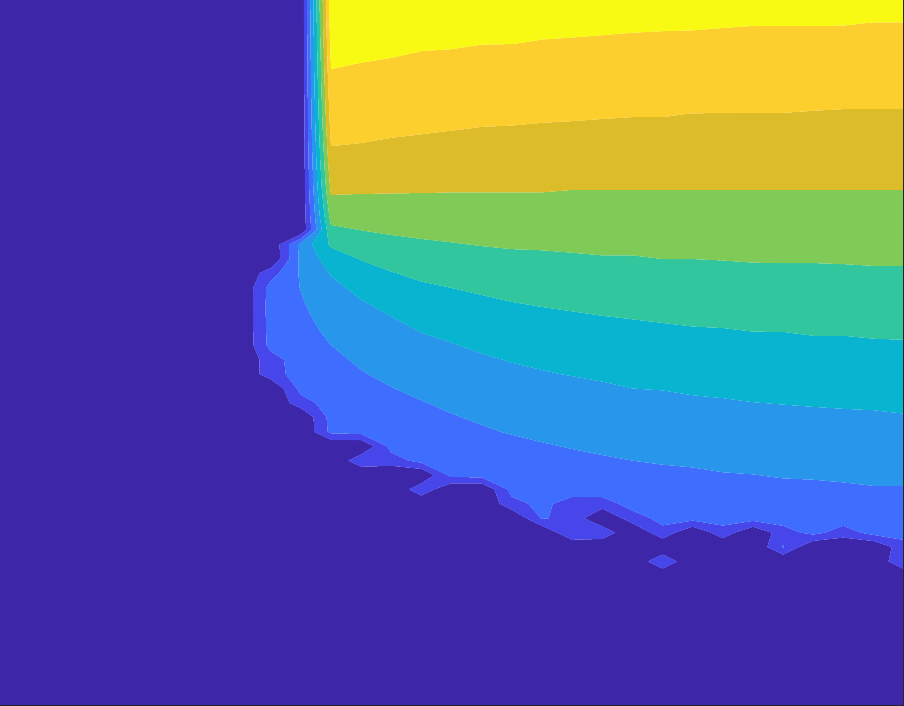};
\end{axis}

\end{tikzpicture}%

%% file: Figures/fig04.tex
%
%
\definecolor{mycolor1}{rgb}{0.00000,0.44700,0.74100}%
\definecolor{mycolor2}{rgb}{0.85000,0.32500,0.09800}%
\definecolor{mycolor3}{rgb}{0.92900,0.69400,0.12500}%
\definecolor{mycolor4}{rgb}{0.49400,0.18400,0.55600}%
\definecolor{mycolor5}{rgb}{0.46600,0.67400,0.18800}%
\definecolor{mycolor6}{rgb}{0.30100,0.74500,0.93300}%
\definecolor{mycolor7}{rgb}{0.63500,0.07800,0.18400}%
\begin{tikzpicture}

\begin{axis}[%
width=72mm,
height=42mm,
at={(0mm,0mm)},
scale only axis,
xmin=-20,
xmax=50,
xticklabel style = {font=\small,yshift=0ex},
xlabel style={font=\small, xshift=2mm},
xlabel={$\beta{}_\text{1}\text{ or }\tilde{\beta}\text{ [dB]}$},
ymode=log,
ymin=0.01,
ymax=100,
yminorticks=true,
yticklabel style = {font=\small,xshift=-2mm},
ylabel style={font=\small, yshift=0mm},
ylabel={Error [m]},
axis background/.style={fill=white},
xmajorgrids,
ymajorgrids,
yminorgrids,
legend style={font=\scriptsize, at={(1,0.25)}, anchor=south east, legend cell align=left, align=left, fill = white, fill opacity=0.8, legend columns = 2}
]
\addplot [color=mycolor1, dashed, line width=1.2pt]
  table[row sep=crcr]{%
-20	0.0433421890599092\\
-15	0.0409092435342488\\
-10	0.037801421804112\\
-5	0.0348363338736413\\
0	0.0335450035910207\\
1	0.0336004140861569\\
2	0.0337646498949301\\
3	0.0340318995801678\\
4	0.0343930324365597\\
5	0.0348363338736413\\
6	0.0353483685945141\\
7	0.0359148584824252\\
8	0.0365214821755558\\
9	0.0371545351004403\\
10	0.037801421804112\\
11	0.0384509796439714\\
12	0.0390936512715404\\
13	0.0397215330794021\\
14	0.0403283297472947\\
15	0.0409092435342487\\
16	0.0414608230461879\\
17	0.0419807913428829\\
18	0.0424678683728699\\
19	0.0429215983527227\\
20	0.0433421890599092\\
21	0.0437303671300582\\
22	0.0440872512857223\\
23	0.0444142438751184\\
24	0.0447129400523509\\
25	0.0449850532753167\\
30	0.0460085935175479\\
35	0.0466182957621976\\
40	0.0469724464846378\\
45	0.047175263102126\\
50	0.0472904905551492\\
};
\addlegendentry{AM: CRB (Bob)}

\addplot [color=mycolor2, dashed, line width=1.2pt]
  table[row sep=crcr]{%
-20	0.0472995845286164\\
-15	0.0471717709250982\\
-10	0.0469231882517487\\
-5	0.046493611695949\\
0	0.0459717659706034\\
1	0.0458821556954536\\
2	0.0458031647618519\\
3	0.0457360560220792\\
4	0.0456812788737285\\
5	0.0456385382121958\\
6	0.0456069496782525\\
7	0.0455852347437129\\
8	0.0455719130577926\\
9	0.045565463320069\\
10	0.0455644398837377\\
11	0.0455675449315497\\
12	0.0455736634781638\\
13	0.0455818711976829\\
14	0.045591424745419\\
15	0.0456017424151925\\
16	0.0456123807589964\\
17	0.0456230107993552\\
18	0.0456333959185312\\
19	0.0456433724307414\\
20	0.0456528331561712\\
21	0.0456617139163984\\
22	0.0456699826665992\\
23	0.0456776308980409\\
24	0.0456846669336506\\
25	0.0456911107661832\\
30	0.0457155434888316\\
35	0.0457301708928989\\
40	0.0457386705021742\\
45	0.0457435355244658\\
50	0.045746298070968\\
};
\addlegendentry{AN: CRB (Bob)}

\addplot [color=mycolor3, line width=1.2pt, mark=o, mark options={solid, mycolor3}]
  table[row sep=crcr]{%
-20	0\\
-15	0\\
-10	0\\
-5	0\\
0	0\\
1	7.60797555849217\\
2	7.58975100559607\\
3	7.54602227187252\\
4	7.52778239841285\\
5	7.50225638669068\\
6	7.45845920913807\\
7	7.44018322844043\\
8	7.44018321438054\\
9	7.39633268488812\\
10	7.39633270599927\\
11	7.37804269836247\\
12	7.35244813388224\\
13	7.35244821806761\\
14	7.33414208377387\\
15	7.3341416920896\\
16	7.30852941339494\\
17	7.29020210899397\\
18	7.29020174709608\\
19	7.29020128328576\\
20	7.29020128276385\\
21	7.29020359856911\\
22	7.29020121168354\\
23	7.27187825554666\\
24	7.27187824137964\\
25	7.24622929953291\\
30	7.24622917341362\\
35	7.24622904587158\\
40	7.24623350667702\\
45	7.24622920432935\\
50	7.24622931056847\\
};
\addlegendentry{AM: Bias (Eve)}

\addplot [color=mycolor4, line width=1.2pt, mark=o, mark options={solid, mycolor4}]
  table[row sep=crcr]{%
-20	0\\
-15	0\\
-10	0\\
-5	0\\
0	0\\
1	0\\
2	0\\
3	0\\
4	0\\
5	0\\
6	0\\
7	0\\
8	0\\
9	0\\
10	0\\
11	0\\
12	0\\
13	0\\
14	0\\
15	0\\
16	0\\
17	0\\
18	0\\
19	0\\
20	0\\
21	36.407478767612\\
22	36.407206655576\\
23	36.4064051109594\\
24	36.4052510099609\\
25	36.409872746898\\
30	36.4086691204969\\
35	36.4052584357711\\
40	36.4070844658009\\
45	36.4085381425237\\
50	36.4092939786089\\
};
\addlegendentry{AN: Bias (Eve)}

\addplot [color=mycolor5, line width=1.2pt, mark=x, mark options={solid, mycolor5}]
  table[row sep=crcr]{%
-20	0.0190399733047829\\
-15	0.0190399733047829\\
-10	0.0190399733047829\\
-5	0.0190399733047829\\
0	0.0190399733047829\\
1	5.92564494543224\\
2	5.91460747237582\\
3	5.88647984023826\\
4	5.87541093800499\\
5	5.85828334819609\\
6	5.8300214983312\\
7	5.81888519266866\\
8	5.81888518615331\\
9	5.7905209584801\\
10	5.79052097144069\\
11	5.77935321302247\\
12	5.76208862454768\\
13	5.76208867685567\\
14	5.75088793088197\\
15	5.75088765491179\\
16	5.73358769694831\\
17	5.72235036610462\\
18	5.72235012491337\\
19	5.72234981542863\\
20	5.72234981514042\\
21	5.7223513708738\\
22	5.72234976857833\\
23	5.71111505568014\\
24	5.71111504943807\\
25	5.69374452155729\\
30	5.69374443581444\\
35	5.69374440128758\\
40	5.69374731081232\\
45	5.69374445924733\\
50	5.69374453128946\\
};
\addlegendentry{AM: MCRB (Eve)}

\addplot [color=mycolor6, line width=1.2pt, mark=x, mark options={solid, mycolor6}]
  table[row sep=crcr]{%
-20	0.0190399733047829\\
-15	0.0190399733047829\\
-10	0.0190399733047829\\
-5	0.0190399733047829\\
0	0.0190399733047829\\
1	0.0190399733047829\\
2	0.0190399733047829\\
3	0.0190399733047829\\
4	0.0190399733047829\\
5	0.0190399733047829\\
6	0.0190399733047829\\
7	0.0190399733047829\\
8	0.0190399733047829\\
9	0.0190399733047829\\
10	0.0190399733047829\\
11	0.0190399733047829\\
12	0.0190399733047829\\
13	0.0190399733047829\\
14	0.0190399733047829\\
15	0.0190399733047829\\
16	0.0190399733047829\\
17	0.0190399733047829\\
18	0.0190399733047829\\
19	0.0190399733047829\\
20	0.0190399733047829\\
21	12.3347072403175\\
22	12.3327532710284\\
23	12.3270276175881\\
24	12.3187628534138\\
25	12.3518035399474\\
30	12.3432109105323\\
35	12.3187971339505\\
40	12.3318693217884\\
45	12.3422676143073\\
50	12.3476700538244\\
};
\addlegendentry{AN: MCRB (Eve)}

\addplot [color=mycolor7, line width=1.2pt, mark=+, mark options={solid, mycolor7}]
  table[row sep=crcr]{%
-20	0.0190399733047829\\
-15	0.0190399733047829\\
-10	0.0190399733047829\\
-5	0.0190399733047829\\
0	0.0190399733047829\\
1	9.64336871212238\\
2	9.62220878381001\\
3	9.57042825777026\\
4	9.54923879313336\\
5	9.51857839592618\\
6	9.46666596249073\\
7	9.4454090095784\\
8	9.44540899448956\\
9	9.39339502821752\\
10	9.39339505282988\\
11	9.37210956080502\\
12	9.34131462260908\\
13	9.3413147211361\\
14	9.3199974301792\\
15	9.31999695166601\\
16	9.28916735047098\\
17	9.26781206663352\\
18	9.26781163303716\\
19	9.26781107710755\\
20	9.26781107651906\\
21	9.2678138587442\\
22	9.26781099185674\\
23	9.24646140654405\\
24	9.24646139154696\\
25	9.21556106475199\\
30	9.21556091260855\\
35	9.21556079098969\\
40	9.21556609615879\\
45	9.21556095139548\\
50	9.21556107944222\\
};
\addlegendentry{AM: LB (Eve)}

\addplot [color=mycolor1, line width=1.2pt, mark=+, mark options={solid, mycolor1}]
  table[row sep=crcr]{%
-20	0.0190399733047829\\
-15	0.0190399733047829\\
-10	0.0190399733047829\\
-5	0.0190399733047829\\
0	0.0190399733047829\\
1	0.0190399733047829\\
2	0.0190399733047829\\
3	0.0190399733047829\\
4	0.0190399733047829\\
5	0.0190399733047829\\
6	0.0190399733047829\\
7	0.0190399733047829\\
8	0.0190399733047829\\
9	0.0190399733047829\\
10	0.0190399733047829\\
11	0.0190399733047829\\
12	0.0190399733047829\\
13	0.0190399733047829\\
14	0.0190399733047829\\
15	0.0190399733047829\\
16	0.0190399733047829\\
17	0.0190399733047829\\
18	0.0190399733047829\\
19	0.0190399733047829\\
20	0.0190399733047829\\
21	38.4402069832936\\
22	38.4393223107\\
23	38.4367264863707\\
24	38.43298348212\\
25	38.4479633288262\\
30	38.4440638162475\\
35	38.4330015039511\\
40	38.4389229852974\\
45	38.4436369187253\\
50	38.4460874963745\\
};
\addlegendentry{AN: LB (Eve)}

\end{axis}

\end{tikzpicture}%

%% file: Figures/fig05a.tex
%
%
\definecolor{mycolor1}{rgb}{0.00000,0.44700,0.74100}%
\definecolor{mycolor2}{rgb}{0.85000,0.32500,0.09800}%
\definecolor{mycolor3}{rgb}{0.92900,0.69400,0.12500}%
\definecolor{mycolor4}{rgb}{0.49400,0.18400,0.55600}%
\definecolor{mycolor5}{rgb}{0.46600,0.67400,0.18800}%
\definecolor{mycolor6}{rgb}{0.30100,0.74500,0.93300}%
\definecolor{mycolor7}{rgb}{0.63500,0.07800,0.18400}%
\begin{tikzpicture}

\begin{axis}[%
width=72mm,
height=42mm,
at={(0in,0in)},
scale only axis,
xmin=-70,
xmax=20,
xticklabel style = {font=\small,yshift=0ex},
xlabel style={font=\small, xshift=2mm},
xlabel={$\text{Transmit power }P\text{ [dBm]}$},
ymode=log,
ymin=0.1,
ymax=1000,
yminorticks=true,
yticklabel style = {font=\small,xshift=-2mm},
ylabel style={font=\small, yshift=0mm},
ylabel={Error [m]},
axis background/.style={fill=white},
xmajorgrids,
ymajorgrids,
yminorgrids,
legend style={font=\scriptsize, at={(1,1)}, anchor=north east, legend cell align=left, align=left, fill = white, fill opacity=0.9, legend columns = 2}
]
\addplot [color=mycolor1, dashed, line width=1.2pt, mark=o]
  table[row sep=crcr]{%
-70	378.01421804112\\
-65	212.573016314037\\
-60	119.538591693745\\
-55	67.2214900644487\\
-50	37.801421804112\\
-45	21.2573016314037\\
-40	11.9538591693745\\
-35	6.72214900644488\\
-40	11.9538591693745\\
-39	10.6538881990237\\
-38	9.49528785215193\\
-37	8.46268420608038\\
-36	7.54237523779876\\
-35	6.72214900644488\\
-34	5.99112160826882\\
-33	5.33959275384294\\
-32	4.75891705111467\\
-31	4.24138928630662\\
-30	3.7801421804112\\
-25	2.12573016314037\\
-20	1.19538591693745\\
-15	0.672214900644487\\
-10	0.37801421804112\\
-5	0.212573016314037\\
0	0.119538591693745\\
5	0.0672214900644488\\
10	0.037801421804112\\
15	0.0212573016314037\\
20	0.0119538591693745\\
};
\addlegendentry{AM: CRB (Bob)}

\addplot [color=mycolor2, dashed, line width=1.2pt]
  table[row sep=crcr]{%
-70	455.644398837377\\
-65	256.22767505777\\
-60	144.087410342429\\
-55	81.0263052752069\\
-50	45.5644398837377\\
-45	25.622767505777\\
-40	14.4087410342429\\
-35	8.1026305275207\\
-40	14.4087410342429\\
-39	12.8418039640951\\
-38	11.4452698303294\\
-37	10.200607473475\\
-36	9.09130098026867\\
-35	8.1026305275207\\
-34	7.22147705900395\\
-33	6.43614820354859\\
-32	5.7362231243805\\
-31	5.1124142409486\\
-30	4.55644398837376\\
-25	2.5622767505777\\
-20	1.44087410342429\\
-15	0.81026305275207\\
-10	0.455644398837376\\
-5	0.25622767505777\\
0	0.144087410342429\\
5	0.0810263052752069\\
10	0.0455644398837377\\
15	0.025622767505777\\
20	0.0144087410342429\\
};
\addlegendentry{AN: CRB (Bob)}

\addplot [color=mycolor3, line width=1.2pt, mark=o, mark options={solid, mycolor3}]
  table[row sep=crcr]{%
-70	7.39633268896663\\
-65	7.39633268896695\\
-60	7.39633268896676\\
-55	7.3963326889669\\
-50	7.39633268896679\\
-45	7.39633268896686\\
-40	7.39633268896681\\
-35	7.39633268896738\\
-40	7.39633268896681\\
-39	7.39633268896709\\
-38	7.39633268896717\\
-37	7.39633268896705\\
-36	7.39633268896717\\
-35	7.39633268896738\\
-34	7.39633268896784\\
-33	7.39633268896771\\
-32	7.39633268896845\\
-31	7.39633268896853\\
-30	7.3963326889692\\
-25	7.39633268897426\\
-20	7.39633268899037\\
-15	7.39633268904121\\
-10	7.3963326892031\\
-5	7.39633268971966\\
0	7.39633269136848\\
5	7.39633435735461\\
10	7.39633270599936\\
15	7.39633303880541\\
20	7.39633406756444\\
};
\addlegendentry{AM: Bias (Eve)}

\addplot [color=mycolor4, line width=1.2pt, mark=o, mark options={solid, mycolor4}]
  table[row sep=crcr]{%
-70	0.23983029234418\\
-65	0.239830292341927\\
-60	0.239830292342552\\
-55	0.239830292342265\\
-50	0.239830292343293\\
-45	0.239830292344988\\
-40	0.239830292348198\\
-35	0.239830292363188\\
-40	0.239830292348198\\
-39	0.23983029235129\\
-38	0.239830292353991\\
-37	0.239830292355167\\
-36	0.239830292358637\\
-35	0.239830292363188\\
-34	0.239830292368094\\
-33	0.239830292375571\\
-32	0\\
-31	0\\
-30	0\\
-25	0\\
-20	0\\
-15	0\\
-10	0\\
-5	0\\
0	0\\
5	0\\
10	0\\
15	0\\
20	0\\
};
\addlegendentry{AN: Bias (Eve)}

\addplot [color=mycolor5, line width=1.2pt, mark=x, mark options={solid, mycolor5}]
  table[row sep=crcr]{%
-70	144.804457146336\\
-65	81.5701859735948\\
-60	46.1195187770042\\
-55	26.3732147311365\\
-50	15.5845414594139\\
-45	9.98656982168604\\
-40	7.38002198695683\\
-35	6.33640978459345\\
-40	7.38002198695683\\
-39	7.08230379707683\\
-38	6.83658467484264\\
-37	6.63492114381609\\
-36	6.47025640828191\\
-35	6.33640978459345\\
-34	6.22804205760608\\
-33	6.14059977282762\\
-32	6.0702443100025\\
-31	6.01377243706253\\
-30	5.96853447093662\\
-25	5.84738746568268\\
-20	5.80855159666938\\
-15	5.79621646384909\\
-10	5.7923102855899\\
-5	5.79107449549033\\
0	5.79068365032837\\
5	5.79056116188638\\
10	5.79052097144074\\
15	5.79050882890889\\
20	5.79050558303712\\
};
\addlegendentry{AM: MCRB (Eve)}

\addplot [color=mycolor6, line width=1.2pt, mark=x, mark options={solid, mycolor6}]
  table[row sep=crcr]{%
-70	192.705055715765\\
-65	108.366201855871\\
-60	60.9391233449709\\
-55	34.2691738254542\\
-50	19.2720154529251\\
-45	10.8393049445224\\
-40	6.09868528243608\\
-35	3.43539770339141\\
-40	6.09868528243608\\
-39	5.43657099725508\\
-38	4.84659641796466\\
-37	4.32093282178097\\
-36	3.85260480492978\\
-35	3.43539770339141\\
-34	3.06377510014631\\
-33	2.73280532108449\\
-32	2.39699062332738\\
-31	2.13632014173831\\
-30	1.90399733047829\\
-25	1.07069638198005\\
-20	0.602096822319172\\
-15	0.338583924955862\\
-10	0.190399733047829\\
-5	0.107069638198005\\
0	0.0602096822319172\\
5	0.0338583924955862\\
10	0.0190399733047829\\
15	0.0107069638198005\\
20	0.00602096822319173\\
};
\addlegendentry{AN: MCRB (Eve)}

\addplot [color=mycolor7, line width=1.2pt, mark=+, mark options={solid, mycolor7}]
  table[row sep=crcr]{%
-70	144.993229313272\\
-65	81.9048287771406\\
-60	46.7088401640238\\
-55	27.3907318723052\\
-50	17.2506136049148\\
-45	12.4272810400864\\
-40	10.4484669580683\\
-35	9.73939557694311\\
-40	10.4484669580683\\
-39	10.240349814336\\
-38	10.0719723620638\\
-37	9.93619221988687\\
-36	9.82700133483238\\
-35	9.73939557694311\\
-34	9.66924222042267\\
-33	9.61315259506177\\
-32	9.56836471028467\\
-31	9.53263846847647\\
-30	9.50416440181214\\
-25	9.42855648653664\\
-20	9.40452066281944\\
-15	9.39690707322227\\
-10	9.39449816083461\\
-5	9.39373626781879\\
0	9.39349532493557\\
5	9.39342112839253\\
10	9.39339505282998\\
15	9.3933878296702\\
20	9.39338663881129\\
};
\addlegendentry{AM: LB (Eve)}

\addplot [color=mycolor1, line width=1.2pt, mark=+, mark options={solid, mycolor1}]
  table[row sep=crcr]{%
-70	192.705204955614\\
-65	108.366467245345\\
-60	60.9395952778052\\
-55	34.2700130325087\\
-50	19.2735076772992\\
-45	10.8419578605284\\
-40	6.10339911388157\\
-35	3.44375898541143\\
-40	6.10339911388157\\
-39	5.44185839372216\\
-38	4.85252670345651\\
-37	4.327583508088\\
-36	3.8600624803364\\
-35	3.44375898541143\\
-34	3.07314764263189\\
-33	2.74330885831119\\
-32	2.39699062332738\\
-31	2.13632014173831\\
-30	1.90399733047829\\
-25	1.07069638198005\\
-20	0.602096822319172\\
-15	0.338583924955862\\
-10	0.190399733047829\\
-5	0.107069638198005\\
0	0.0602096822319172\\
5	0.0338583924955862\\
10	0.0190399733047829\\
15	0.0107069638198005\\
20	0.00602096822319173\\
};
\addlegendentry{AN: LB (Eve)}

\end{axis}
\end{tikzpicture}%

%% file: Figures/fig05b.tex
%
%
\definecolor{mycolor1}{rgb}{0.00000,0.44700,0.74100}%
\definecolor{mycolor2}{rgb}{0.85000,0.32500,0.09800}%
\definecolor{mycolor3}{rgb}{0.92900,0.69400,0.12500}%
\definecolor{mycolor4}{rgb}{0.49400,0.18400,0.55600}%
\definecolor{mycolor5}{rgb}{0.46600,0.67400,0.18800}%
\definecolor{mycolor6}{rgb}{0.30100,0.74500,0.93300}%
\definecolor{mycolor7}{rgb}{0.63500,0.07800,0.18400}%
\begin{tikzpicture}

\begin{axis}[%
width=72mm,
height=42mm,
at={(0in,0in)},
scale only axis,
xmin=-70,
xmax=20,
xticklabel style = {font=\small,yshift=0ex},
xlabel style={font=\small, xshift=2mm},
xlabel={$\text{Transmit power }P\text{ [dBm]}$},
ymode=log,
ymin=0.1,
ymax=1000,
yminorticks=true,
yticklabel style = {font=\small,xshift=-2mm},
ylabel style={font=\small, yshift=0mm},
ylabel={Error [m]},
axis background/.style={fill=white},
xmajorgrids,
ymajorgrids,
yminorgrids,
legend style={font=\scriptsize, at={(0.0,0.0)}, anchor=south west, legend cell align=left, align=left, fill = white, fill opacity=0.9, legend columns = 2}
]
\addplot [color=mycolor1, dashed, line width=1.2pt, mark=o]
  table[row sep=crcr]{%
-70	460.085935175479\\
-65	258.72533448802\\
-60	145.491947456309\\
-55	81.8161345371057\\
-50	46.0085935175479\\
-45	25.872533448802\\
-40	14.5491947456309\\
-35	8.18161345371058\\
-30	4.60085935175479\\
-25	2.5872533448802\\
-20	1.45491947456309\\
-15	0.818161345371058\\
-10	0.460085935175479\\
-5	0.25872533448802\\
0	0.14549194745631\\
5	0.0818161345371057\\
10	0.0460085935175479\\
15	0.025872533448802\\
20	0.014549194745631\\
};
\addlegendentry{AM: CRB (Bob)}

\addplot [color=mycolor2, dashed, line width=1.2pt]
  table[row sep=crcr]{%
-70	471.261782149462\\
-65	265.009975085494\\
-60	149.026060578238\\
-55	83.8035123934636\\
-50	47.1261782149462\\
-45	26.5009975085494\\
-40	14.9026060578238\\
-35	8.38035123934636\\
-30	4.71261782149462\\
-25	2.65009975085494\\
-20	1.49026060578238\\
-15	0.838035123934636\\
-10	0.471261782149462\\
-5	0.265009975085494\\
0	0.149026060578238\\
5	0.0838035123934636\\
10	0.0471261782149462\\
15	0.0265009975085494\\
20	0.0149026060578238\\
};
\addlegendentry{AN: CRB (Bob)}

\addplot [color=mycolor3, line width=1.2pt, mark=o, mark options={solid, mycolor3}]
  table[row sep=crcr]{%
-70	7.24622916697784\\
-65	7.24622916697786\\
-60	7.24622916697782\\
-55	7.24622916697773\\
-50	7.24622916697776\\
-45	7.24622916697786\\
-40	7.24622916697767\\
-35	7.24622916697758\\
-30	7.24622916697648\\
-25	7.24622916697315\\
-20	7.24622916696348\\
-15	7.24622916693226\\
-10	7.24622916683421\\
-5	7.2462291665291\\
0	7.24622916562515\\
5	7.24622916369017\\
10	7.24622917341418\\
15	7.24622907802245\\
20	7.2462261505308\\
};
\addlegendentry{AM: Bias (Eve)}

\addplot [color=mycolor4, line width=1.2pt, mark=o, mark options={solid, mycolor4}]
  table[row sep=crcr]{%
-70	39.7595637964648\\
-65	39.7595633388451\\
-60	39.7595632219816\\
-55	39.7595634809552\\
-50	39.7595634052271\\
-45	39.7595639778172\\
-40	39.7595630523485\\
-35	39.7595625950858\\
-30	39.7595602050424\\
-25	39.7595536379792\\
-20	39.7595347267884\\
-15	39.7594732002224\\
-10	39.7592772018754\\
-5	39.7586605013557\\
0	39.7567071826776\\
5	39.7505364114316\\
10	39.7310654723262\\
15	39.6699130274898\\
20	39.4807187674862\\
};
\addlegendentry{AN: Bias (Eve)}

\addplot [color=mycolor5, line width=1.2pt, mark=x, mark options={solid, mycolor5}]
  table[row sep=crcr]{%
-70	145.401548462695\\
-65	81.9007391529069\\
-60	46.296195228147\\
-55	26.4565638442905\\
-50	15.6048208040988\\
-45	9.958495876008\\
-40	7.31626313773427\\
-35	6.2525056640528\\
-30	5.87617509080659\\
-25	5.75204709435171\\
-20	5.71223305174638\\
-15	5.69958486602859\\
-10	5.69557931299963\\
-5	5.69431205938573\\
0	5.69391125952\\
5	5.69378450910875\\
10	5.69374443581481\\
15	5.69373169374346\\
20	5.69372574832003\\
};
\addlegendentry{AM: MCRB (Eve)}

\addplot [color=mycolor6, line width=1.2pt, mark=x, mark options={solid, mycolor6}]
  table[row sep=crcr]{%
-70	93.6868941072611\\
-65	58.0504428060347\\
-60	40.7417500894213\\
-55	33.4537326894368\\
-50	30.7921407036577\\
-45	29.9012050870936\\
-40	29.6138884443607\\
-35	29.5224458397545\\
-30	29.4934624562234\\
-25	29.4842660078018\\
-20	29.4812733990032\\
-15	29.4800654039896\\
-10	29.4788585777417\\
-5	29.4758620964995\\
0	29.4666548391758\\
5	29.4376424341796\\
10	29.3461505178543\\
15	29.0591181712986\\
20	28.1734681714183\\
};
\addlegendentry{AN: MCRB (Eve)}

\addplot [color=mycolor7, line width=1.2pt, mark=+, mark options={solid, mycolor7}]
  table[row sep=crcr]{%
-70	145.581997968464\\
-65	82.2206720413599\\
-60	46.859849868977\\
-55	27.4309607485297\\
-50	17.2051814715334\\
-45	12.3158222321057\\
-40	10.2973561383943\\
-35	9.5708758334527\\
-30	9.32937676579505\\
-25	9.2516962183122\\
-20	9.22699537106289\\
-15	9.21917050415935\\
-10	9.21669468133496\\
-5	9.21591161869095\\
0	9.21566397781762\\
5	9.21558566391093\\
10	9.21556091260922\\
15	9.21555296504322\\
20	9.21554698982752\\
};
\addlegendentry{AM: LB (Eve)}

\addplot [color=mycolor1, line width=1.2pt, mark=+, mark options={solid, mycolor1}]
  table[row sep=crcr]{%
-70	101.774540238462\\
-65	70.3610459478278\\
-60	56.9272612019199\\
-55	51.9612848094848\\
-50	50.2889531735164\\
-45	49.7484169935745\\
-40	49.5762568464856\\
-35	49.5216884416695\\
-30	49.5044134936982\\
-25	49.4989297612601\\
-20	49.4971320676003\\
-15	49.4963631530915\\
-10	49.4954869322655\\
-5	49.4932069196766\\
0	49.486154764947\\
5	49.4639256133109\\
10	49.3938671677247\\
15	49.1745294690463\\
20	48.5022830720705\\
};
\addlegendentry{AN: LB (Eve)}

\end{axis}

\end{tikzpicture}%